\documentclass{aa}  
\usepackage{graphicx}
\usepackage{txfonts}
\usepackage{graphicx}	
\usepackage{amsmath, amssymb}	
\usepackage{bm}		
\usepackage{caption}
\usepackage{subcaption}
\usepackage{xcolor}
\usepackage{multicol}        
\usepackage{chemformula} 
\usepackage{placeins}

\newcommand{\repeatcaption}[2]{%
  \renewcommand{\thefigure}{\ref{#1}}%
  \captionsetup{list=no}%
  \caption{#2}%
  \addtocounter{figure}{-1}
}
\usepackage[]{hyperref}	
\hypersetup{colorlinks=true,linkcolor=blue,citecolor=blue,filecolor=blue,urlcolor=blue,}
\begin{document} 

   \title{Pattern Finding in mm-Wave Spectra of Massive Young Stellar Objects}

   \subtitle{ }

   \author{Yenifer Angarita
          \inst{1,2}\fnmsep\thanks{(YA) y.angarita@astro.ru.nl, yenifer.angarita@chalmers.se}
          \and
          Germ\'an Chaparro\inst{3}
          \and
          Stuart L. Lumsden\inst{4}
          \and
          Catherine Walsh\inst{4}
          \and
          Adam Avison\inst{5,6}
          \and
          Naomi Asabre Frimpong\inst{7}
          \and
          Gary A. Fuller\inst{6}
          }

   \institute{Department of Astrophysics/IMAPP, Radboud University, PO Box 9010, 6500 GL Nijmegen, The Netherlands
         \and
             Department of Space, Earth \& Environment, Chalmers University of Technology, 412 93 Gothenburg, Sweden
         \and
             FACom, Instituto de F\'isica - FCEN, Universidad de Antioquia, Calle 70 No. 52-21, Medell\'in, Colombia
         \and
             School of Physics and Astronomy, University of Leeds, Woodhouse Lane, LS2 9JT, Leeds, UK
         \and
             SKA Observatory, Jodrell Bank, Lower Withington, Macclesfield SK11 9FT, UK
         \and
             Jodrell Bank Centre for Astrophysics, Department of Physics and Astronomy, School Of Natural Science, The University of Manchester, Manchester, M13 9PL, UK
         \and
             Ghana Space Science and Technology Institute, Ghana
             }

   \date{Received 30 August 2024; accepted 23 December 2024}

 
%
\abstract
    {Massive stars (M$_\ast>8$M$_\odot$) play a pivotal role in shaping their galactic surroundings due to their high luminosity and intense ionizing radiation. However, the precise mechanisms governing the formation of massive stars remain elusive. Complex organic molecules (COMs) offer an avenue for studying star formation across the low- to high-mass spectrum because COMs are found in every young stellar object (YSO) phase and offer insight into the structure and temperature. We aim to unveil patterns in the evolution of COM chemistry in 41 massive young stellar objects (MYSOs) sourced from diverse catalogues, using Atacama Large Millimeter/Submillimeter Array Band 6 spectra. Previous line analysis of these sources has shown the presence of methanol, methyl acetylene, and methyl cyanide with diverse excitation temperatures (a few tens to hundreds of Kelvin) and column densities (spanning two to four orders of magnitude in range), indicating a possible evolutionary path across sources. However, this analysis usually involves manual line extraction and rotational diagram fitting. Here, we improve upon this process by directly retrieving the physicochemical state of MYSOs from their dimensionally-reduced spectra. We use a Locally Linear Embedding to find a lower-dimensional projection for the physicochemical parameters obtained from individual line analysis. We identify clusters of similar MYSOs in this embedded space using a Gaussian Mixture Model. We find three groups of MYSOs corresponding to distinct physicochemical conditions: i) cold, COM-poor sources, ii) warm, medium-COM-abundance sources, and iii) hot, COM-rich sources. We then use principal component analysis (PCA) on the source spectra to obtain eigenspectra components, finding further evidence for an evolutionary path across MYSO groups. Finally, we find that the physicochemical state of MYSOs in our sample can be derived directly from the spectra by training a simple random forest model on the first few spectral PCA components. Our results highlight the effectiveness of dimensionality reduction in obtaining clear physical insights directly from MYSO spectra.}

   \keywords{methods: data analysis --
             astrochemistry -- 
             stars: formation --
             ISM: clouds --
             stars: protostars --
             radio lines: stars
               }

   \maketitle
%
\section{Introduction}

\begin{figure*}
    	\includegraphics[width=\linewidth]{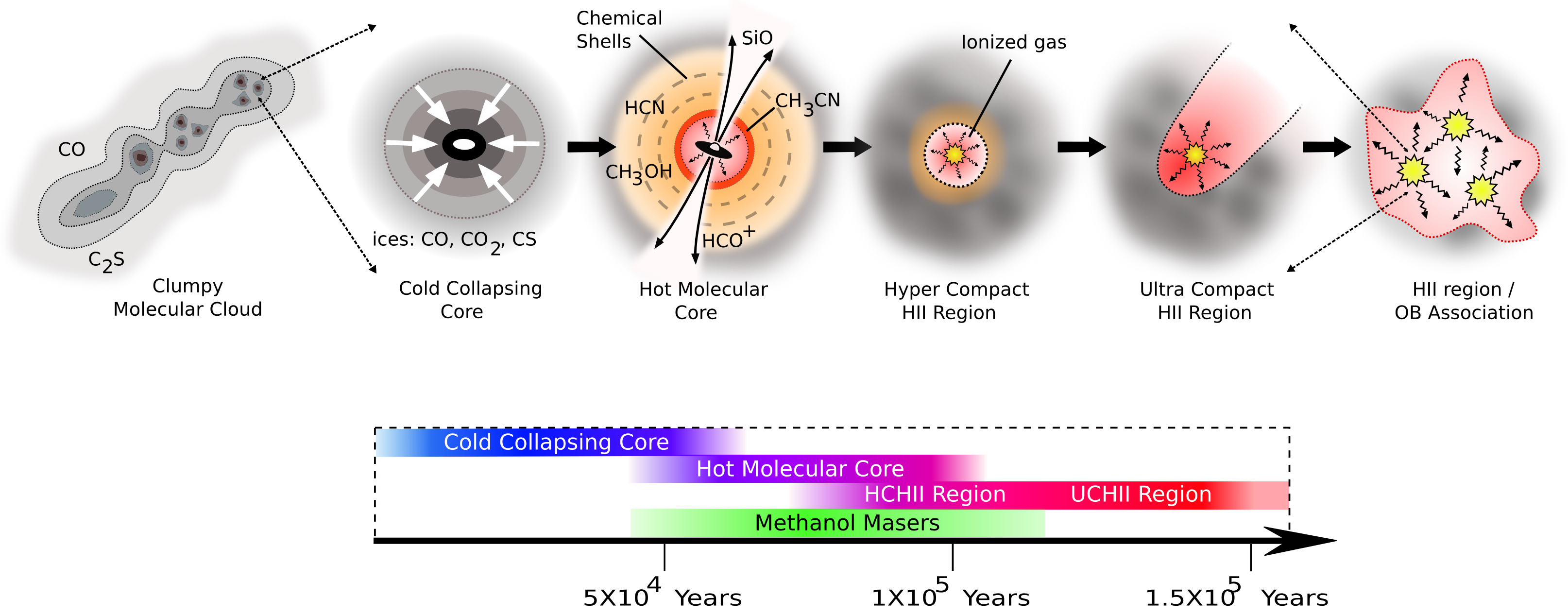}
        \caption{Evolutionary stages of high-mass star formation. Cartoon credit: Dr. Cormac R. Purcell.}
        \label{fig:MYSO_formation_phases}
    \end{figure*}

    Giant molecular clouds (GMC) are dense and cold agglomerations of gas (mainly molecular hydrogen) and dust in the interstellar medium (ISM) and are the sites of star formation \citep{carroll_ostlie_2017}. GMCs fragment hierarchically via gravitational collapse into smaller and denser clumps with masses between $10^3{-}10^5$~M$_\odot$, and temperatures of $10{-}20$~K \citep{van_den_Ancker1999, Williams_2000}, for example, see a cartoon with the evolutionary stages of high-mass star formation in Fig.~\ref{fig:MYSO_formation_phases}. The clumps continue to collapse under the influence of their gravity into denser protostellar cores where the formation of massive young stellar objects (MYSOs) and multiple stellar systems begins \citep[see the review from][and references therein]{Williams_2000}. The massive cold collapsing cores can become infrared dark clouds (IRDCs), hot molecular cores (HMCs), and \mbox{H\,{\sc ii}} regions \citep{Kurtz_2000, Kurtz_2005, Menten_2005, Hoare_2007}. Late star-forming stages (from ${\sim}10^4$ years, Fig.~\ref{fig:MYSO_formation_phases}) are traced by the emergence of outflows, jets, bright infrared (IR) sources, masers of methanol and water, and compact \mbox{H\,{\sc ii}} regions, \citep[see e.g.][]{Hoare_2007, Kurtz_2005, Menten_1991, Shepherd&Churchwell2_1996}. However, clumps and cold cores (the early stages on the left side of Fig.~\ref{fig:MYSO_formation_phases}) do not have those signatures already present, which makes them more challenging to study.
    %

    Complex organic molecules (COMs) are observed in different evolutionary stages of high-mass star formation \citep{Herbst_vanDishoeck_2009}. COMs offer unique insight into the physical and chemical structure of the core at the earliest stages of massive star formation. Therefore, studying their formation and evolution is fundamental for understanding the high-mass star formation process. In particular, IRDCs are filamentary structures containing cold ($T{<}100$~K), massive cores ($10$~M$_\odot$ to $2\times10^3$~M$_\odot$) where stars are forming and the first generation of molecular species--the precursors of COMs--is formed in the ice mantles via surface chemistry. Due to their size, density, and mass similarities with HMCs, IRDCs are considered an earlier evolutionary phase of HMCs \citep{Rathborne_2006}. On the other hand, HMCs are compact (${<}0.1$~pc), warm ($T{>}100$~K), and dense (\mbox{$n > 10^6$~cm$^{-3}$}) sources with an elevated abundance of complex molecules at small spatial scales proposed to be driven by ice sublimation \citep[e.g.~see][and references therein]{Millar_1996}.

    COMs such as \ch{CH3OH} (methanol), \ch{CH3CN} (methyl cyanide), and \ch{CH3CCH} (methyl acetylene) are essential in tracing the evolution of hot cores and cold extended structures \citep{Isokoski_2013, Bisschop_2007, Oberg_2014}. For instance, \cite{Fayolle_2015} studied MYSOs with weak hot organic emission lines, finding that N-bearing molecules are generally concentrated in the cores while O- and C-bearing molecules are present both in the cores and the envelopes. Nevertheless, larger surveys of MYSOs with ice detections are needed to quantify the impact of the initial conditions on COM formation. Observations at high sensitivity, spatial, and spectral resolution from the Atacama Large Millimeter/Sub-Millimeter Array (ALMA) \citep[e.g.~see][]{Avison_2023} are detecting COM emission from regions that are usually optically thick at most frequencies. \cite{Frimpong_2021} performed an extended and exhaustive analysis of a large spectral sample of MYSOs observed by \cite{Avison_2023}. Among all molecules studied by \cite{Frimpong_2021}, species such as \ch{CH3OH}, \ch{CH3CN}, and \ch{CH3CCH} efficiently describe the physical and chemical conditions of the sources, showing two distinct evolutionary stages and hinting at a possible third intermediate stage. These results thus open new possibilities for spectral MYSO classification. However, the analysis and interpretation of large samples of chemically rich spectra demands considerable effort and time due to the wide variety of species present and the inherent heterogeneity of the physical conditions. 
    

    Classification methods for young stellar object (YSO) spectra have been successfully implemented along with simulations and observations \citep{Singh_1998, Ronen_1999, Yip_2004, Ward&Lumsden_2016}. For instance, \cite{Ward&Lumsden_2016} demonstrated the potential of dimensionality reduction techniques, such as locally linear embedding (LLE) and principal component analysis (PCA), to efficiently classify large spectral samples based on the presence/absence of emission lines. In this work, we build on this basic idea and show that dimensionality reduction techniques are also robust and helpful in studying chemical evolution in large samples of MYSO spectra. We find that the first eight PCA components of the original spectra retain sufficient physicochemical information to classify the sources comparably to methods based on molecular excitation temperature and column density obtained from manual line extraction and rotational diagram fitting. Thus, reduced-dimensionality spectra can convey information about a source's chemical evolutionary stage, which is particularly useful in distinguishing between COM-rich and COM-poor sources. This blind, PCA-based approach to source classification is more efficient than the traditional rotational diagram fitting method in terms of human pre-processing time, as it requires little individual source or line inspection and extraction.
       
    
    Section~\ref{sec:obs_data} presents the list of MYSOs, between HMCs and IRDCs, selected for our study and observed by ALMA. Section~\ref{sec:Dimen_Reduc} details the dimensionality reduction and classification methods. Our results are outlined in Sect.~\ref{sec:Results}. We discuss the implications of our results in Sect.~\ref{sec:Discu} and summarise our work in Sect.~\ref{sec:Summary}.

\section{Data and observations}
\label{sec:obs_data}
        
    Our sample consists of 43 MYSOs (Table~\ref{tab:sources}), of which 17 are from the Red MSX Source (RMS) database \citep{Lumsden_2013}, and 26 objects are from the Spitzer Dark Cloud (SDC) catalogue \citep[][also see \citealt{Traficante_2015}]{Peretto_Fuller_2009}. The MYSO sample represents a mixture of sources in different evolutionary stages, comprising hot cores in an advanced evolutionary stage and younger, denser, and colder objects such as IRDCs. 
    %
    \begin{table*}
        \caption{\label{tab:sources}Sample of sources and their physical parameters.}
        \centering
        \resizebox{\textwidth}{!}{
        \begin{tabular}{lcccccccccc} 
            \hline\hline
                (1) & (2) & (3) & (4) & (5) & (6) & (7) & (8) & (9) & (10) & (11)\\
                Name & $\alpha$ (J2000) & $\delta$ (J2000) & $D$ & $R_\text{gc}$ & $L$ & $V_\text{lsr}$ & $T$ & $M_\text{gas}$ & $N_\text{gas}$ & Radio Flux\\ 
                & (hh:mm:ss) & (dd:mm:ss) & (kpc) & (kpc) & ($10^4L_{\odot}$) & (km~s$^{-1}$) & (K) & ($M_{\odot}$) & ($10^{24}$~cm$^{-2}$) & (mJy) \\
            \hline
                G013.6562-00.5997 & 18:17:24.248 & -17:22:12.840 & 4.10 & 4.62 & 1.40 & 49.0 & 500.0 & 12.18 & 2.350 & 239.17 \\
                G017.6380+00.1566 & 18:22:26.000 & -13:30:11.987 & 2.20 & 6.44 & 10.00 & 23.0 & 200.0 & 3.65 & 2.450 & 97.93 \\
                G023.3891+00.1851 & 18:33:14.320 & -08:23:57.372 & 4.50 & 4.72 & 2.40 & 76.2 & 350.0 & 11.68 & 1.870 & 132.62 \\
                G029.8620-00.0444 & 18:45:59.570 & -02:45:06.619 & 4.90 & 4.90 & 2.80 & 101.2 & 270.0 & 8.09 & 1.090 & 59.45 \\
                G030.1981-00.1691 & 18:47:03.050 & -02:30:36.814 & 4.90 & 4.93 & 3.00 & 104.8 & 200.0 & 14.17 & 1.920 & 76.56 \\
                G034.7569+00.0247 & 18:54:40.740 & 01:38:06.372 & 4.60 & 5.40 & 1.20 & 77.5 & 350.0 & 6.96 & 1.070 & 75.58 \\
                G034.8211+00.3519 & 18:53:37.929 & 01:50:30.097 & 3.50 & 5.97 & 2.40 & 57.2 & 220.0 & 8.04 & 2.130 & 93.91 \\
                G050.2213-00.6063 & 19:25:57.520 & 15:03:00.330 & 3.30 & 6.87 & 1.30 & 39.8 & 15.0 & 34.74 & 10.400 & 21.24 \\
                G326.6618+00.5207 & 15:45:02.860 & -54:09:03.090 & 1.80 & 7.07 & 1.40 & -40.0 & 15.0 & 32.26 & 32.400 & 66.30 \\
                G327.1192+00.5103 & 15:47:32.721 & -53:52:38.620 & 4.90 & 5.13 & 3.70 & -83.8 & 300.0 & 40.98 & 5.550 & 335.46 \\
                G332.0939-00.4206 & 16:16:16.460 & -51:18:25.315 & 3.60 & 5.58 & 9.30 & -56.0 & 500.0 & 13.64 & 3.420 & 347.33 \\
                G332.9636-00.6800 & 16:21:22.950 & -50:52:58.625 & 3.20 & 5.83 & 2.20 & -48.0 & 300.0 & 2.42 & 0.769 & 46.52 \\
                G332.9868-00.4871 & 16:20:37.810 & -50:43:49.555 & 3.60 & 5.54 & 1.80 & -55.2 & 220.0 & 4.94 & 1.240 & 54.52 \\
                G333.0682-00.4461 & 16:20:48.980 & -50:38:40.492 & 3.60 & 5.54 & 4.00 & -52.5 & 300.0 & 25.67 & 6.440 & 389.28 \\
                G338.9196+00.5495 & 16:40:34.009 & -45:42:07.251 & 4.20 & 4.82 & 3.20 & -62.5 & 260.0 & 7.47 & 1.380 & 71.90 \\
                G339.6221-00.1209 & 16:46:05.988 & -45:36:43.646 & 2.80 & 5.96 & 1.90 & -32.8 & 200.0 & 7.75 & 3.210 & 128.34 \\
                G345.5043+00.3480 & 17:04:22.887 & -40:44:22.757 & 2.00 & 6.58 & 10.00 & -17.8 & 300.0 & 14.90 & 12.100 & 731.88 \\
                SDC18.816-0.447\_1 & 18:26:59.058 & -12:44:46.518 & 4.30 & 4.64 & 0.46 & 62.0 & 30.0 & 33.31 & 5.850 & 29.60 \\
                SDC20.775-0.076\_1 & 18:29:16.417 & -10:52:11.569 & 4.00 & 4.97 & 0.65 & 57.0 & 25.0 & 89.33 & 18.100 & 73.38 \\
                SDC20.775-0.076\_3.0 & 18:29:12.060 & -10:50:36.012 & 4.00 & 4.97 & 0.58 & 57.0 & 170.0 & 5.46 & 1.110 & 37.47 \\
                SDC20.775-0.076\_3.1 & 18:29:12.263 & -10:50:33.641 & 4.00 & 4.97 & 0.58 & 58.0 & 200.0 & 7.37 & 1.500 & 59.73 \\
                SDC22.985-0.412\_1 & 18:34:40.282 & -09:00:38.256 & 4.60 & 4.63 & 0.31 & 78.5 & 250.0 & 42.85 & 6.580 & 330.35 \\
                SDC23.21-0.371\_1 & 18:34:55.201 & -08:49:14.832 & 3.80 & 5.23 & 0.98 & 78.5 & 300.0 & 70.99 & 16.000 & 966.25 \\
                SDC24.381-0.21\_3\tablefootmark{*} & 18:36:40.780 & -07:39:15.701 & 3.60 & 5.43 & 0.55 & ... & ... & ... & ... & ... \\
                SDC24.462+0.219\_2 & 18:35:11.300 & -07:26:31.465 & 6.30 & 3.80 & 0.73 & 119.2 & 200.0 & 38.80 & 3.180 & 126.84 \\
                SDC25.426-0.175\_6 & 18:37:30.379 & -06:41:18.751 & 4.00 & 5.18 & 1.02 & -14.0 & 200.0 & 25.09 & 5.100 & 203.48 \\
                SDC28.147-0.006\_1\tablefootmark{a} & 18:42:42.606 & -04:15:35.483 & 4.50 & 5.00 & 0.50 & 98.0 & 300.0 & 11.80 & 1.890 & 114.50 \\
                SDC28.277-0.352\_1 & 18:44:21.974 & -04:17:39.734 & 3.10 & 5.95 & 0.54 & 85.2 & 220.0 & 5.36 & 1.810 & 79.87 \\
                SDC29.844-0.009\_4 & 18:46:12.950 & -02:39:01.273 & 5.40 & 4.67 & 0.27 & 102.2 & 300.0 & 6.35 & 0.708 & 42.80 \\
                SDC30.172-0.157\_2.0 & 18:47:07.900 & -02:30:04.485 & 4.20 & 5.31 & 0.68 & 106.2 & 10.0 & 8.84 & 1.630 & 1.78 \\
                SDC30.172-0.157\_2.1\tablefootmark{*} & 18:47:07.913 & -02:30:01.091 & 4.20 & 5.31 & 0.68 & ... & ... & ... & ... & ... \\
                SDC33.107-0.065\_2 & 18:52:07.964 & 00:08:11.694 & 4.50 & 5.33 & 1.92 & 78.2 & 200.0 & 31.86 & 5.110 & 204.13 \\
                SDC35.063-0.726\_1\tablefootmark{a} & 18:58:06.149 & 01:37:07.632 & 2.30 & 6.75 & 0.52 & 36.0 & 380.0 & 2.00 & 1.230 & 94.63 \\
                SDC37.846-0.392\_1\tablefootmark{b} & 19:01:53.563 & 04:12:49.227 & 4.10 & 5.83 & 4.35 & 62.0 & ... & ... & ... & ... \\
                SDC42.401-0.309\_2 & 19:09:49.859 & 08:19:45.508 & 4.50 & 6.00 & 0.61 & 66.5 & 250.0 & 0.16 & 0.026 & 1.29 \\
                SDC43.186-0.549\_2.1\tablefootmark{a} & 19:12:09.028 & 08:52:14.193 & 4.20 & 6.15 & 1.44 & 60.0 & 120.0 & 15.70 & 2.890 & 67.96 \\
                SDC43.186-0.549\_2.2 & 19:12:09.200 & 08:52:14.961 & 4.20 & 6.15 & 1.44 & 60.5 & 70.0 & 34.36 & 6.330 & 83.77 \\
                SDC43.311-0.21\_1 & 19:11:17.212 & 09:07:31.565 & 4.30 & 6.13 & 1.08 & 59.0 & 350.0 & 17.52 & 3.080 & 217.86 \\
                SDC43.877-0.755\_1 & 19:14:26.364 & 09:22:35.843 & 3.20 & 6.58 & 0.87 & 54.2 & 360.0 & 4.62 & 1.470 & 106.69 \\
                SDC45.787-0.335\_1 & 19:16:31.082 & 11:16:11.990 & 4.50 & 6.26 & 0.64 & 62.0 & 250.0 & 21.39 & 3.430 & 172.33 \\
                SDC45.927-0.375\_2.0\tablefootmark{a} & 19:16:56.230 & 11:21:58.048 & 4.20 & 6.34 & 1.01 & 63.5 & 55.0 & 7.80 & 1.440 & 14.61 \\
                SDC45.927-0.375\_2.1\tablefootmark{c} & 19:16:56.106 & 11:21:53.010 & 4.20 & 6.34 & 1.01 & 62.2 & ... & ... & ... & ... \\
                SDC45.927-0.375\_2.2 & 19:16:56.165 & 11:21:50.471 & 4.20 & 6.34 & 1.01 & 60.8 & 60.0 & 11.76 & 2.170 & 24.22\\
            \hline
        \end{tabular}
        }
        \tablefoot{This table is adapted from \cite{Avison_2023} and \cite{Frimpong_2021}. Distances and luminosities are taken from \citet{Lumsden_2013} and \citet{Peretto_Fuller_2009}. The rest of the parameters are from \cite{Avison_2023} and \cite{Frimpong_2021}. Columns: (1) Name of the sources, (2) and (3) RA and DEC coordinates, respectively, (4) kinetic distances, derived from CO emission lines and rotation curves from \citet{Reid_2009}, (5) Galactocentric distance calculated with the kinematic distance, the Galactic centre distance from the Sun (8.5~kpc adopted by the International Astronomical Union), and the Galactic longitude of the sources, (6) bolometric luminosity, (7) local standard rest frame velocity, and (8) gas temperatures assumed from the excitation temperature of \ch{CH3CN} (where the temperature of \ch{CH3CN} was not available the temperature of \ch{CH3OH} was used instead), (9) and (10) the \ch{H2} column density ($N_\text{gas}$) and mass ($M_\text{gas}$), respectively, and (11) the radio flux measured from the continuum maps at $1.24{-}1.33$~mm.\\ 
        \tablefoottext{a}{Spectrum normalized by the brightest line in the \ch{CH3OH} $5{-}4$ ladder at average rest frequency \mbox{$\nu_\text{lsr} \approx 241.75$~GHz}.\\}
        \tablefoottext{b}{Spectrum normalized by the brightest line in the \ch{CH3CCH} $14{-}13$ ladder between rest frequencies \mbox{$\nu_\text{lsr} \approx 239.13{-}239.25$~GHz}.\\}
        \tablefoottext{c}{Spectrum normalized by the peak intensity of \ch{C^{34}S} at average rest frequency \mbox{$\nu_\text{lsr} \approx 241.01609$~GHz}.\\}
        \tablefoottext{*}{Spectrum not considered for PCA due to a lack of \ch{HC3N}, \ch{H2CO}, \ch{CH3OH}, \ch{CH3CN}, and \ch{CH3CCH} emission.}
        }
    \end{table*}

\begin{figure}
         \centering
         \begin{subfigure}[b]{0.98\columnwidth}
             \centering
             \includegraphics[width=\textwidth]{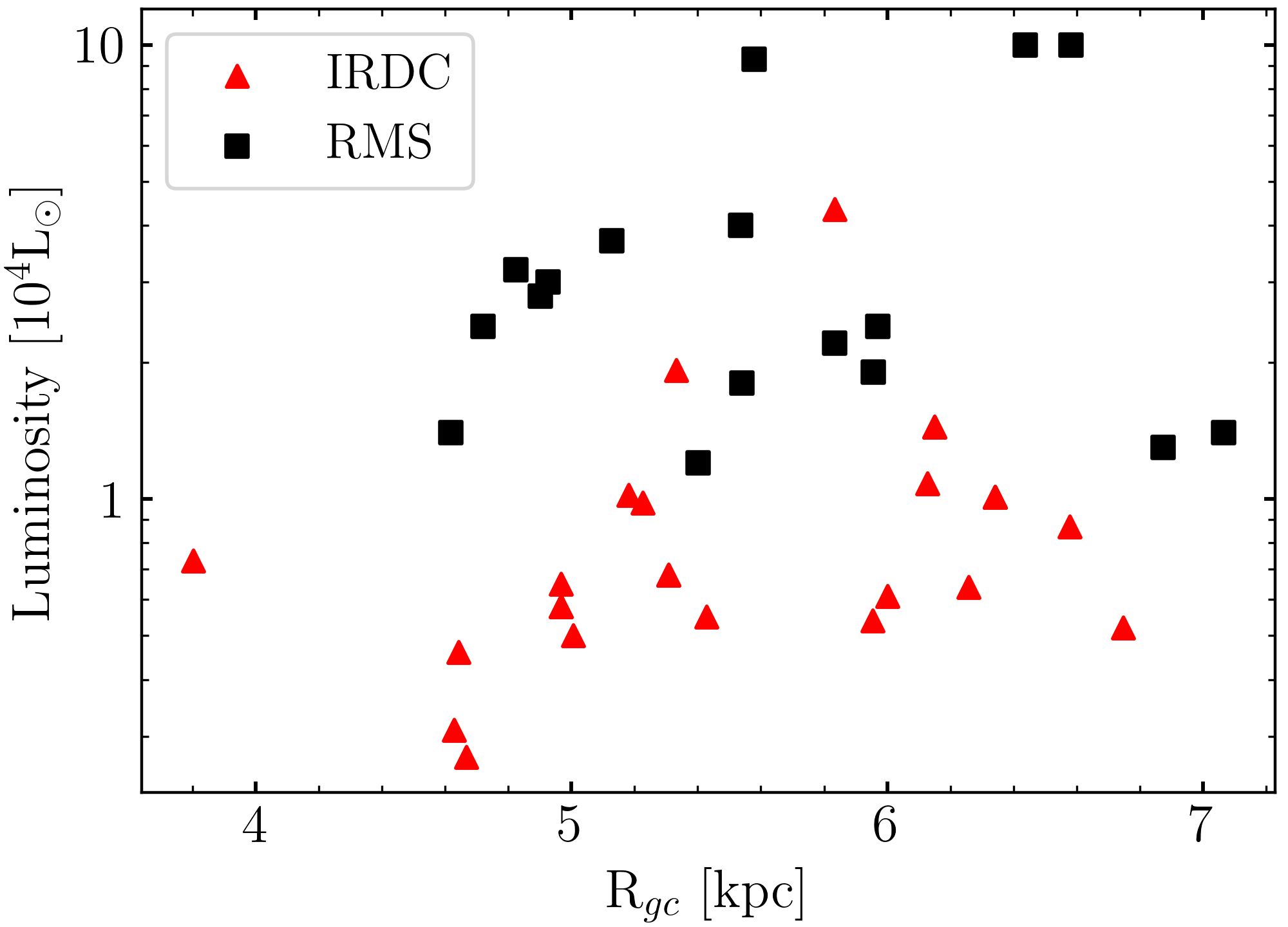}
         \end{subfigure}
         \begin{subfigure}[b]{0.98\columnwidth}
             \centering
             \includegraphics[width=\textwidth]{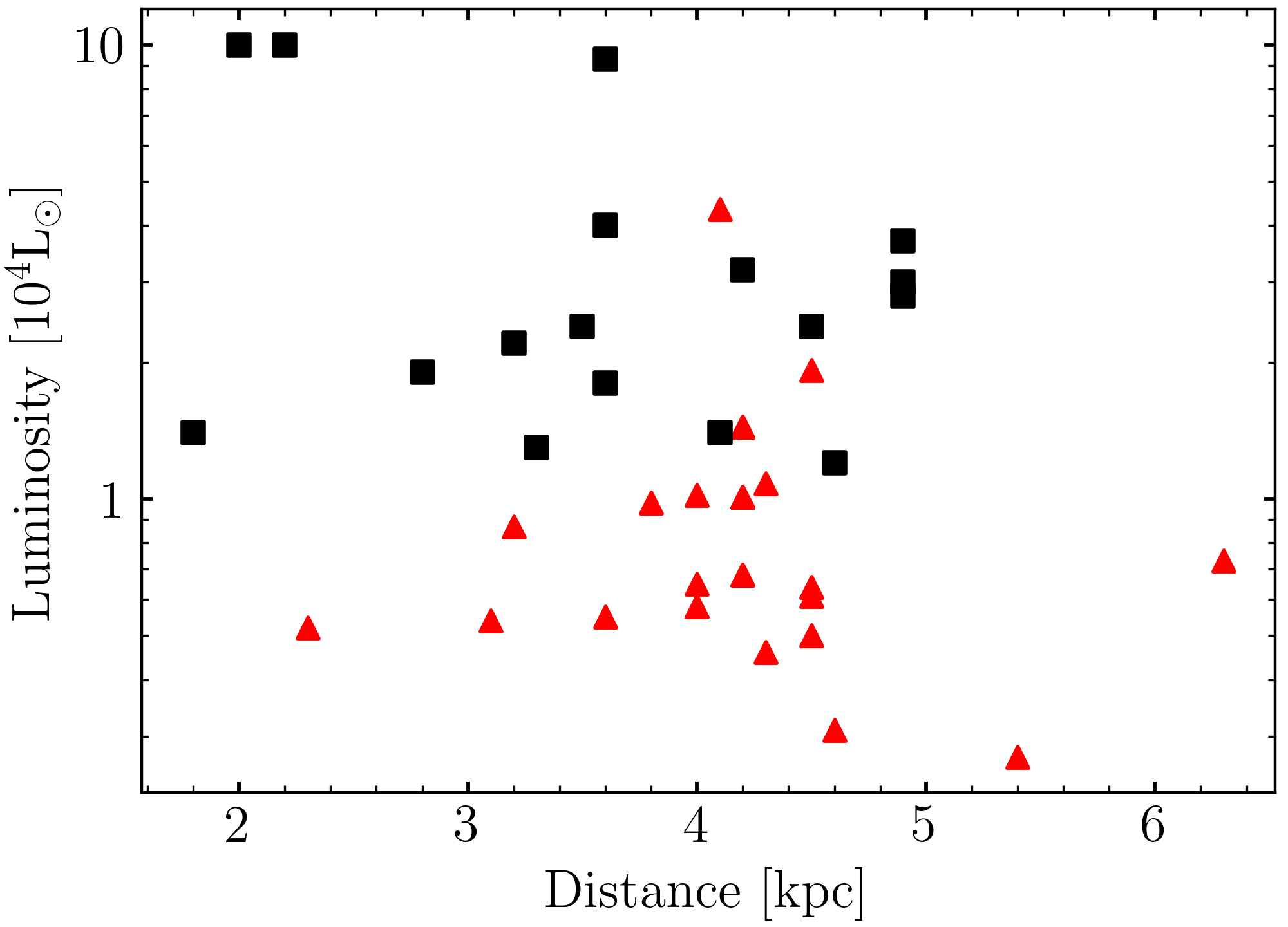}
         \end{subfigure}
            \caption{Luminosity as a function of Galactocentric radius (top) and the kinematic distance (bottom) of the sample. The black and red squares are the RMS and IRDC sources, respectively.}
            \label{fig:Rgc-Lum_Sources}
    \end{figure}

    The objects from our sample are located near the Galactic plane and have continuum observations available in the mid and far infrared. Figure~\ref{fig:Rgc-Lum_Sources} shows the bolometric luminosity as a function of the Galactocentric radius (top) and the distance from the Sun (bottom). The majority of the sources lie between ${\sim}4$ and ${\sim}7$~kpc from the Galactic centre and between ${\sim}2$ and ${\sim}6$~kpc from the Sun, with no clear distance preference among YSOs from different catalogues. On the other hand, The luminosities of RMS sources are typically higher than those of the IRDCs. The broad range in luminosity can be attributed to the embedded nature of the youngest objects in the sample rather than a distance bias.
    %
    
    \subsection{ALMA observations}
    \label{subsec:Observ} 
        In Cycle 3 at Band 6, ALMA conducted observations with high sensitivity, spatial, and spectral resolution of $38$ fields, from which our sample of $43$ objects (Table~\ref{tab:sources}) was drawn. The wavelength regime is around $1.24{-}1.33$~mm with four unique spectral windows (SPWs) covering a total bandwidth of $1.875$~GHz each (central frequencies are in Table~\ref{tab:SPWs}). The SPW frequency and velocity resolutions are $976.562$~kHz and ${\sim}1.25$~km~s$^{-1}$, respectively. Six epochs of observations were carried out with the $12$~m main arrays of $36{-}39$ antennas in dual polarization mode. The baselines were between $15{-}460$~m. The average resolution is between ${\sim}0\farcs7{-}0\farcs8$ and the largest angular scales between $5\farcs79{-}9\farcs98$. Further details of the ALMA observations are found in \cite{Avison_2023}. 
        %
        \begin{table}
         \caption{SPWs frequencies.}
         \label{tab:SPWs}
         \centering
         \begin{tabular}{c c c}
            \hline
            Spectral  & Central Frequency & Frequency Range\\
             window   &       (GHz)       &      (GHz)     \\
        	\hline
                    0 & 239.789 & ${\sim}238.8{-}240.7$\\
                    1 & 241.844 & ${\sim}240.8{-}242.7$ \\
                    2 & 227.095 & ${\sim}226.1{-}228.0$ \\
                    3 & 225.206 & ${\sim}224.2{-}226.1$ \\ 
        	\hline
         \end{tabular}
        \end{table}
        
    \subsection{Calibration, reduction, and spectra extraction}
    \label{subsec:cal_red}
    
        Calibration and reduction of observations were carried out with the standard ALMA pre-pipeline calibration and imaging with the Common Astronomy Software Applications (CASA) package \citep{McMullin_2007}. The calibration, reduction, and imaging team \citep{Avison_2023} searched for line-free channels to subtract the continuum using the LumberJack code \footnote{\url{https://github.com/adam-avison/LumberJack}}. Then, imaging was performed on the data using the continuum-subtracted data. Due to the complexity of the mm-wave structures within each field of view, only the brightest MYSOs were selected for the spectra extraction. The spectra of 43 sources were extracted by \cite{Frimpong_2021} using CASA from regions delimited by a circular mask of $1\arcsec$ radii (marginally resolved) centred at the source’s peak intensity coordinates. These procedures are explained in detail by \cite{Frimpong_2021}.

        Table~\ref{tab:sources} presents the relevant parameters for each of the 43 MYSOs obtained from \cite{Lumsden_2013}, \cite{Peretto_Fuller_2009}, \cite{Avison_2023}, and \cite{Frimpong_2021}. In particular, there were no \ch{CH3OH}, \ch{CH3CN}, and \ch{CH3CCH} detections for \mbox{SDC24.381-0.21\_3} and \mbox{SDC30.172-0.157\_2.1}. Therefore, they are not considered further in the analysis, leading to a true sample size of 41 MYSOs. Our dimensionality-reduction-based classification scheme (Sects.~\ref{subsec:Dimen_Reduc_LLE},~\ref{subsec:Dimen_Reduc_GMM}, and~\ref{subsec:Result_LLE_GMM}) uses a subsample of 32 objects, excluding sources without excitation temperature and density data.

    \subsection{Post-reduction}
    \label{subsec:post-reduc}
        
        Before attempting any classification scheme, we had to standardize our spectral sample. Our procedure is as follows: 1) shift the spectra to the local standard rest (LSR) frame, 2) normalize all spectra, and 3) interpolate the spectra to a standard frequency grid.

        \subsubsection{Local standard rest frame and Doppler shift correction}
        \label{subsec:post-reduc_LSR_dopplerShift}

            To correct the Doppler shift accurately, we must first determine the proper LSR velocity ($V_\text{lsr}$) for each source. \cite{Peretto_Fuller_2009} and \cite{Lumsden_2013} provide a $V_\text{lsr}$ measurement (also adopted by \citealt{Frimpong_2021}) from observations of molecular lines such as ammonia, \ch{NH3} (1, 1), carbon monosulfide, CS ($2{-}1$), formyl radical, HCO ($3{-}2$), carbon monoxide, CO, and some of its isotopologues \citep[see][for more details]{Peretto_Fuller_2009, Lumsden_2013}. Still, due to the complexity of the high spatial and spectral resolution observations (see Sect.~\ref{subsec:cal_red}), the velocities from the literature need to be re-adjusted for the spectra extracted. 

\begin{figure}
             \centering
                 \begin{subfigure}[b]{0.98\columnwidth}
                     \centering
                     \includegraphics[width=\textwidth]{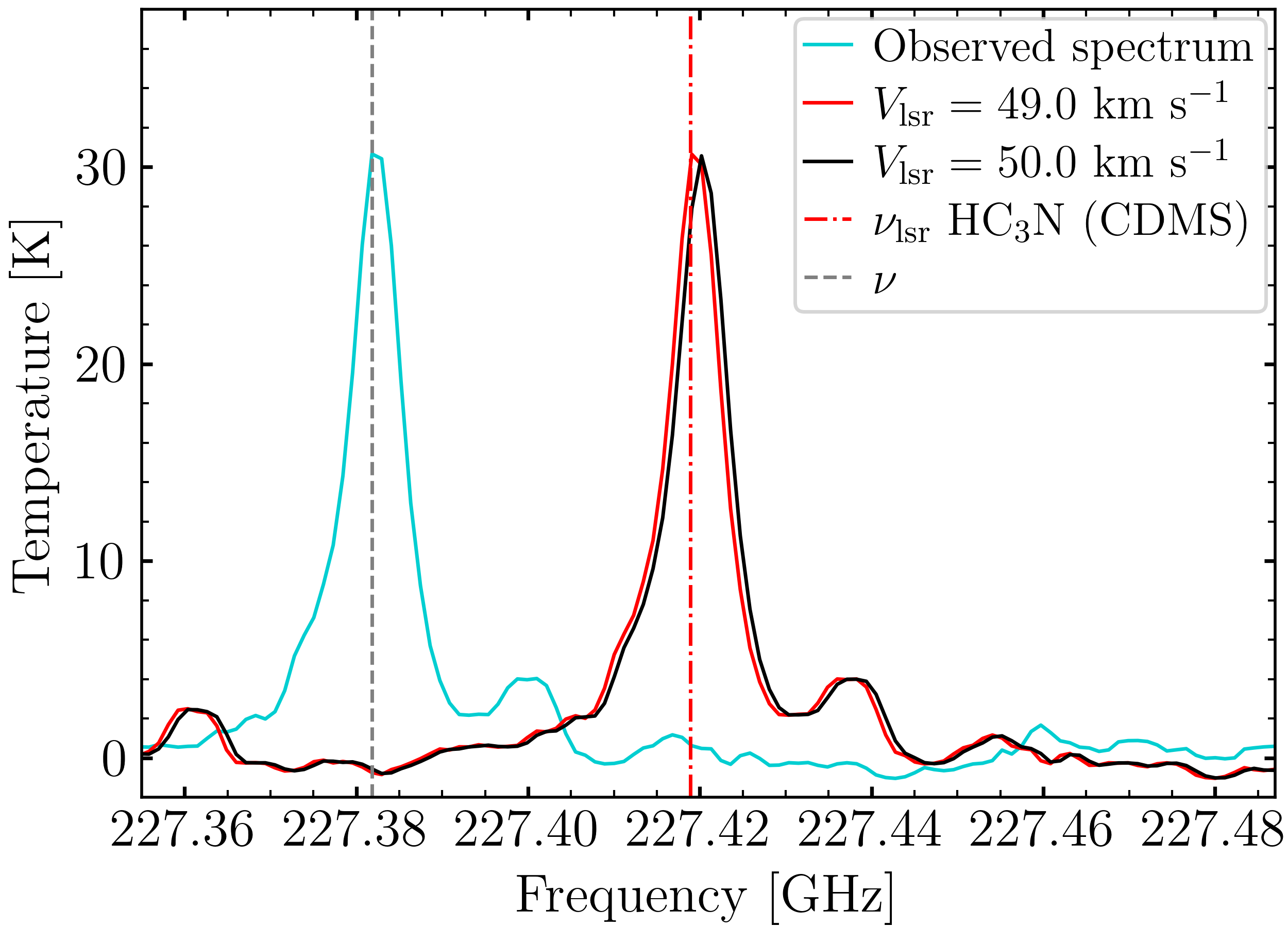}
                 \end{subfigure}
                 \hfill
                 \begin{subfigure}[b]{0.98\columnwidth}
                     \centering
                     \includegraphics[width=\textwidth]{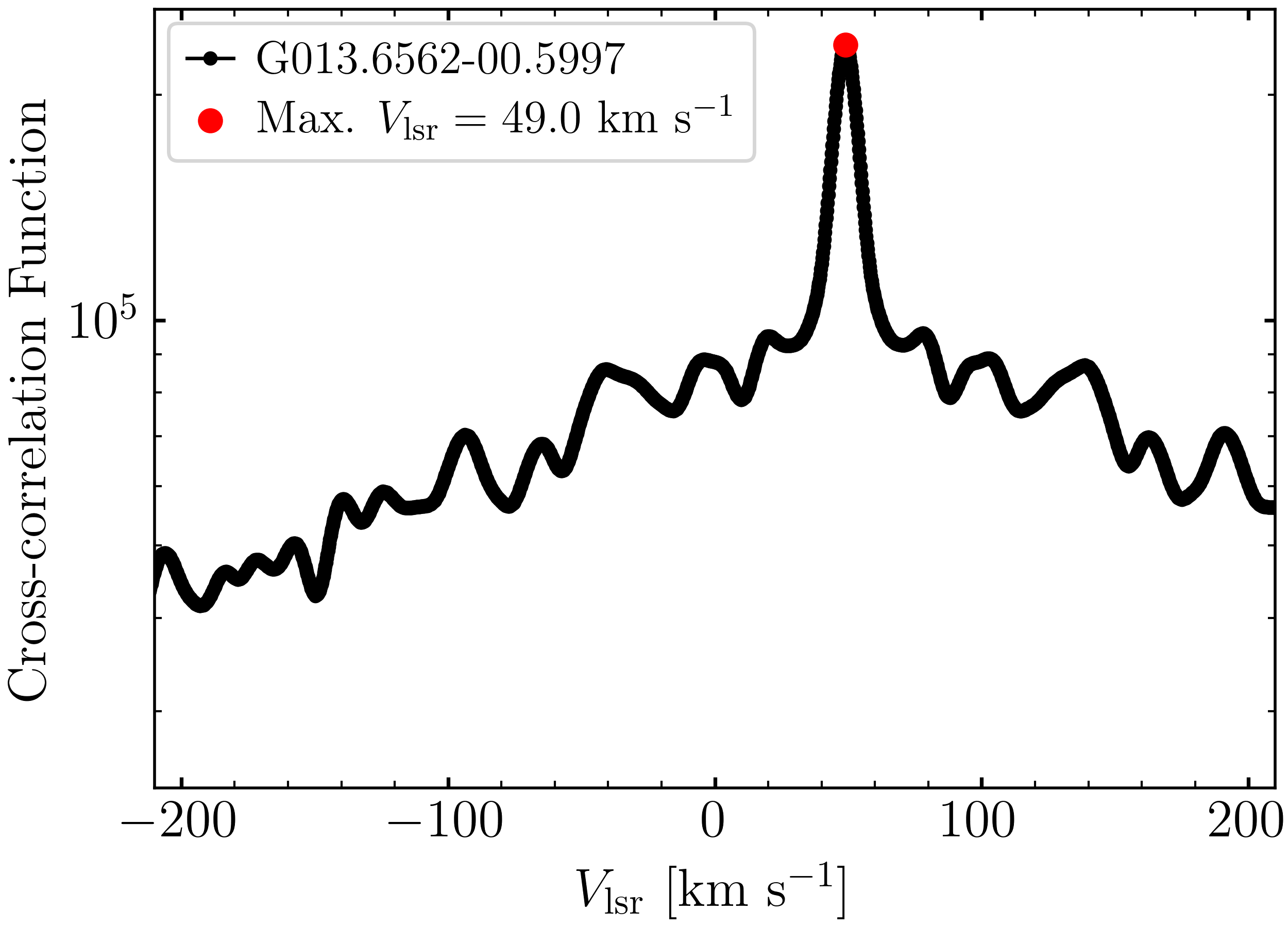}
                 \end{subfigure}
                 \caption{Top: profile of \ch{HC3N} (cyanoacetylene) line emission in \mbox{G013.6562-00.5997} with rest frequency \mbox{$\nu_\text{lsr} = 227.41891$~GHz} (the vertical red dot-dashed line). The turquoise curve is the observed line with central frequency marked by the vertical grey dashed line. The black curve is the line corrected by \mbox{$V_\text{lsr} = 50.0$~km~s$^{-1}$} \citep{Lumsden_2013, Frimpong_2021}. The red curve is the line corrected by \mbox{$V_\text{lsr}=49.0$~km~s$^{-1}$}. Bottom: cross-correlation function of the source \mbox{G013.6562-00.5997} maximized at $V_\text{lsr} = 49.0$~km~s$^{-1}$.}
                 \label{fig:HC3N_LSR_dopplershift}
            \end{figure}

            Firstly, we calculated our first guess of $V_\text{lsr}$ in one source (\mbox{G013.6562-00.5997}) using the observed central frequency $\nu$ of the bright, well detected \ch{HC3N} (cyanoacetylene) line and its central rest frequency \mbox{$\nu_\text{lsr} = 227.41891$~GHz}\footnote{Value from the Cologne Database for Molecular Spectroscopy (CDMS). \url{https://cdms.astro.uni-koeln.de/}} as follows,
            \begin{equation}
            \label{eq:Vlsr}
                V_\text{lsr} = \left(  1 - \frac{\nu}{\nu_\text{lsr}} \right)c\ .
            \end{equation}
            Equation~(\ref{eq:Vlsr}) finds the velocity at radio frequencies, where $c$ is the speed of light. Figure~\ref{fig:HC3N_LSR_dopplershift} (top) shows the observed \ch{HC3N} line emission in \mbox{G013.6562-00.5997} (the turquoise solid line). Secondly, we shift the spectrum of \mbox{G013.6562-00.5997} using the velocity found with Eq.~(\ref{eq:Vlsr}), $V_\text{lsr}=48.89$~km~s$^{-1}$, and the Python package dopplershift\footnote{\url{https://pyastronomy.readthedocs.io/en/latest/pyaslDoc/aslDoc/dopplerShift.html}} from \textsc{PyAstronomy} library \citep{Czesla_PyAstro_2019}. 

            Finally, we use the cross-correlation method to correct the Doppler shift of the 41 spectra. The technique consists of computing the cross-correlation of a test spectrum and a template signal and finding its maximum for deriving the relative velocity between them \citep{Allende_2007}. We use the shifted spectrum of \mbox{G013.6562-00.5997} as the template to find the cross-correlation function and corresponding $V_\text{lsr}$ values of the entire sample with the Python package crosscorrRV\footnote{\url{https://pyastronomy.readthedocs.io/en/latest/pyaslDoc/aslDoc/crosscorr.html}} \citep{Czesla_PyAstro_2019}; for example, see Fig.~\ref{fig:HC3N_LSR_dopplershift}, bottom. 
            
            The original spectrum of \mbox{G013.6562-00.5997} was also included in the cross-correlation process because the initial guess for $V_\text{lsr}$ was based on the central frequency observed at the peak line intensity, without accounting for a suitable model of the observed line. Figure~\ref{fig:HC3N_LSR_dopplershift} (top) demonstrates the difference between the spectrum shifted using $V_\text{lsr}=49.0$~km~s$^{-1}$ (the red solid line) and the one obtained with the typical value from literature (the black solid line), that is, $V_\text{lsr} = 50.0$~km~s$^{-1}$ \citep{Lumsden_2013}. 
            The cross-correlation yielded unexpected results for the source \mbox{SDC37.846-0.392\_1}, likely because we used the same frequency window in the cross-correlation for all spectra. Therefore, we decided to adopt the value from the literature \citep{Peretto_Fuller_2009, Frimpong_2021}. 
            The final $V_\text{lsr}$ value for each source is presented in Table~\ref{tab:sources}.
            %
            %
            %
            \begin{figure*}
             \centering
                \includegraphics[width=\linewidth]{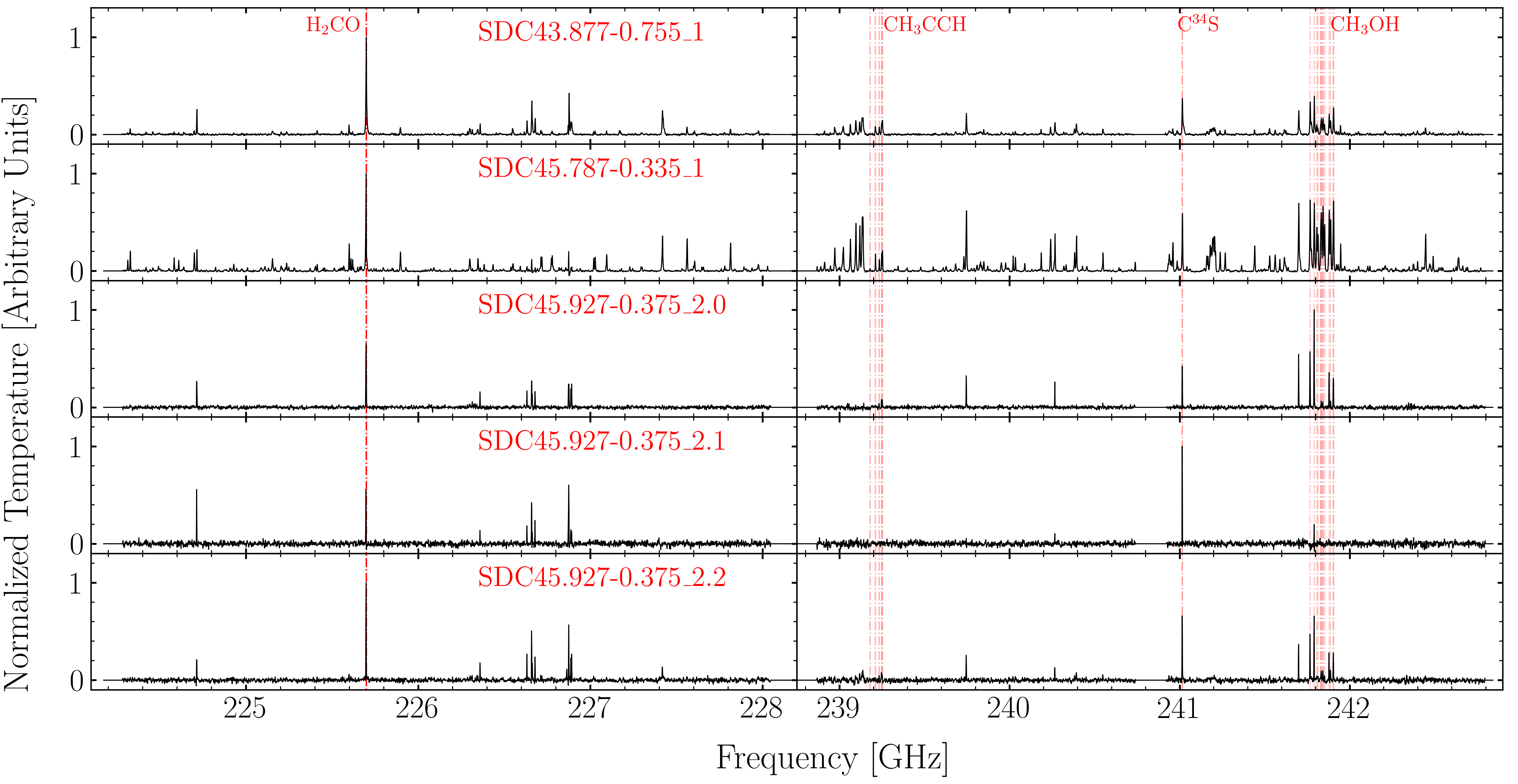}
                \caption{Standardized spectra of five sources from our sample. The remaining spectra are provided in the Appendix~\ref{appendix:sample}. The vertical red lines indicate the emission lines used in the normalization (See Sect.~\ref{subsec:post-reduc_norm} for more details).}
                \label{fig:norm_spectra_sample1}
            \end{figure*}

        \subsubsection{Normalization}
        \label{subsec:post-reduc_norm}
        
            The next step in the post-reduction process is to normalize the spectra to reduce or prevent significant variances due to brightness temperature differences. We normalized the spectra by the most common and brightest molecular line within the sample, \ch{H2CO} (formaldehyde), with \mbox{$\nu_{\text{lsr}} = 225.69778$~GHz}. This molecular species, formed mainly via grain-surface chemistry through hydrogenation, has proven to be one of the most abundant molecules in MYSOs envelopes and predecessor of important COMs such as \ch{CH3OH} \citep[methanol,][]{watanabe_2002, Hidaka_2004, Garrod_2006, Santos_2022}. 
            
            When \ch{H2CO} is not detected nor bright enough for the normalization, we use the brightest line in the \ch{CH3OH} $5{-}4$ ladder (\mbox{$\nu_{\text{lsr}} \approx 241.75$~GHz}), in the \ch{CH3CCH} $14{-}13$ ladder \mbox{($239.13{-}239.25$~GHz)}, or the \ch{C^{34}S} line (\mbox{$241.01609$~GHz}), which is the brightest in \mbox{SDC45.927-0.375\_2.1} (see the marks in Table~\ref{tab:sources}). Figure~\ref{fig:norm_spectra_sample1} shows the final normalized spectra and rest frequency of the above lines (the vertical red dashed lines). 

        \subsubsection{Gaps and edge issues in frequency}
        \label{subsec:post-reduc_gap_issues}

            Due to the shift in the frequency axis, some channels (i.e.~frequencies) cannot be interpolated at the edges of the SPWs because they are outside the given input range. If this is not treated correctly, the spectra will be trimmed in the edges, and the number of channels will not stay constant across all spectra. In cases where this is relevant, we specified an edge-handling parameter in the Doppler shift algorithm (Sect.~\ref{subsec:post-reduc_LSR_dopplerShift}). When this parameter is activated, the algorithm fills the missing frequencies with zeros so that all new arrays remain with the same axis dimension. 
            
            Additionally, the gaps between SPWs were filled with zero values, guaranteeing parity in the frequency axis of the sample. In this way, regardless of the number of interpolations made during the Doppler shift correction, the dimension of the frequency axis will remain the same in the entire spectral sample, and the gaps will never contribute to the emission. However, it should be noted that some information in the spectra edges is lost.
            
            Non-identical frequency channels can appear between sources due to different observational conditions. Therefore, we created a standard, synthetic frequency grid of dimension equal to the normalized spectra. The new grid takes values between the minimum and maximum frequency across the spectra, with an interval equal to the smallest range between adjacent frequencies. The final sample shown in Fig.~\ref{fig:norm_spectra_sample1} was interpolated to the new synthetic grid of frequencies.

    \subsection{Complex organic molecules}
    \label{subsec:COMs_selected}      
        
        As mentioned above, COMs are extremely useful in characterizing hot cores and cold structures \citep{Bisschop_2007, Isokoski_2013, Oberg_2014, Fayolle_2015}. \ch{CH3OH} seems to be always present at early stages of formation ($T{\sim}10$~K) and remains in the hot core with a wide variation in abundance of $10^{-8}{-}10^{-6}$ relative to \ch{H2} \citep{Bisschop_2007, Yusef-Zadeh_2013}. \ch{CH3CN} is efficiently formed in hot cores ($T{>}100$~K) but is unlikely to be detected in cold sources \citep{Bisschop_2007, Fayolle_2015}. On the other hand, \ch{CH3CCH} is usually found in extended cold sources \citep{Bisschop_2007, Isokoski_2013, Fayolle_2015}. Therefore, these chemical species are fundamental to understanding how the chemical and physical conditions of the observed regions are evidenced in the spectra. Since we aim to extract as much information from the overall spectra via dimensionality reduction, we will analyse the species that might or might not be present in the eigenspectra.
        
        \cite{Frimpong_2021} did the first chemical analysis of the spectral sample (Fig.~\ref{fig:norm_spectra_sample1} and Table~\ref{tab:sources}), showing the detection of at least 12 COMs and simple molecules such as \ch{H2CO}, \ch{SO2}, and \ch{CN}. \cite{Frimpong_2021} and \cite{Frimpong_2023subm} extracted enough information from the COM lines to make rotational diagrams for most objects. The rotational temperatures and column densities obtained from this analysis showed some correlations, which we will discuss briefly.
        
        \ch{CH3OH} and \ch{CH3CN} excitation temperatures were found to range from a few tens of Kelvin to more than several hundred Kelvin, whereas \ch{CH3CCH} emission is always below $100$~K in our sample (see \citealp{Frimpong_2021} and \citealp{Frimpong_2023subm}). Most sources have \ch{CH3OH} and \ch{CH3CN} emission above $100$~K, so they are likely hosts to HMCs. Meanwhile, low \ch{CH3OH} and \ch{CH3CN} temperatures (${<}100$~K) may be a sign of emission coming from cold regions, for example, the envelopes. Similarly, the excitation temperature of \ch{CH3CCH}, always below $100$~K, likely originates from cold material surrounding the hot cores. Additionally, COM column densities were found to be divided into two groups of low and high values, correlating with the excitation temperatures as shown in \cite{Frimpong_2021}. We should note that the column density of \ch{CH3CCH} emission measured in \cite{Frimpong_2021} is higher on average than the values reported in the literature from other MYSOs \citep[e.g.][]{Oberg_2014}. 
        The excitation temperatures and column densities of the COMs studied here are from \cite{Frimpong_2021} and \cite{Frimpong_2023subm}. Table~\ref{tab:COMs_param} shows the availability of parameters for each COM and object in the sample.

\section{Dimensionality reduction and classification techniques}
\label{sec:Dimen_Reduc}

\begin{figure}
     \centering
        \includegraphics[width=0.98\columnwidth]{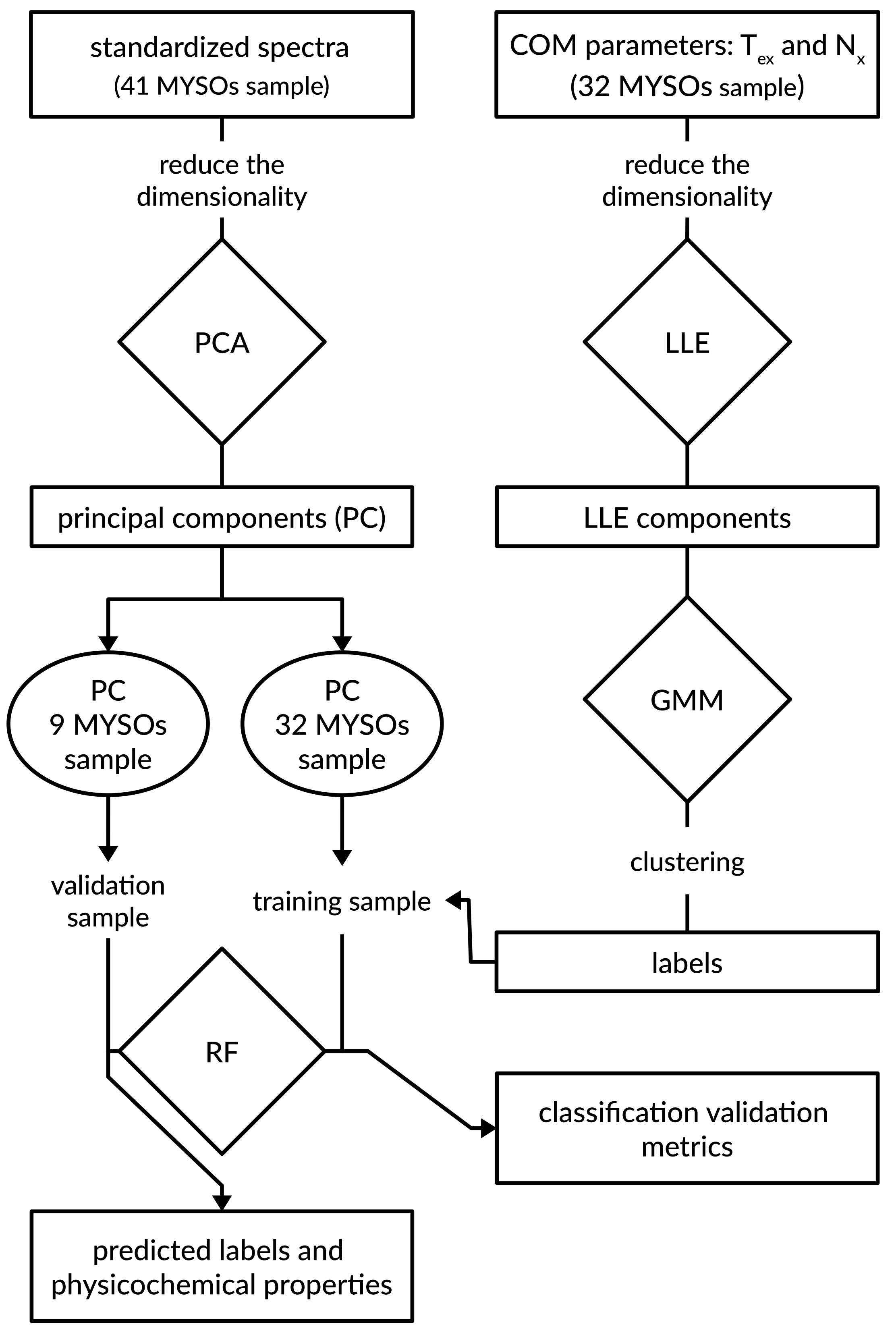}
        \caption{Steps for dimensionality reduction and classification of the MYSOs sample.}
        \label{fig:flowchart}
    \end{figure}

    Dimensionality reduction methods help find patterns related to high-dimensional data and provide a robust classification method \citep[e.g.~see][]{Roweis_2000, Jolliffe_2002}. In this work, we investigate whether the physical information from individual line rotational diagrams can be inferred from a dimensionality reduction analysis of the spectral sample. We start with a PCA of the standardized spectra. To simplify the PCA-to-physical state mapping, we also reduce the dimensionality of the rotational diagram information via low dimensional embedding of the excitation temperature and column density via LLE and Gaussian mixture models (GMMs). We then apply a random forest (RF) supervised classifier to the PCA components to test whether it can learn the physical state of the YSO from these components.
    We outline the details of the methods used in the following sections and summarise the steps in the flowchart of Fig.~\ref{fig:flowchart}. 
    %

    \subsection{Principal component analysis}
    \label{subsec:Dimen_Reduc_PCA}

        PCA \citep{Pearson_1901, Jolliffe_2002} is a linear transformation that aims to help in pattern recognition by searching for the principal components or ``axes'' along which the data characteristics show the greatest variance. At the same time, these principal components should allow for a reconstruction of the original features. PCA has previously been used in the spectral classification of stars \citep{Singh_1998}, galaxies \citep{Ronen_1999}, and quasi-stellar objects \citep{Yip_2004}. In this research, PCA is used to find patterns within several mm-wave spectral features (Sect.~\ref{subsec:COMs_selected}) of 41 MYSOs (See Table~\ref{tab:sources} and Fig.~\ref{fig:norm_spectra_sample1}), that can be correlated to the physical and chemical nature of these type of sources in different evolutionary stages. 

        To implement PCA or any dimensionality reduction technique, each entry (i.e.~each spectrum) of the spectral sample of dimension $N$ must have the same dimension $M$ with the exact correspondence in the frequency axis, that is, a matrix $A$ with dimensions $N\times M$. Therefore, the normalization, gap-handle, and edge-handle (Sects.~\ref{subsec:post-reduc_norm} and \ref{subsec:post-reduc_gap_issues}) ensures no variance contribution from significant differences in observed brightness temperatures and phantom emission.  

    \subsection{Locally linear embedding}
    \label{subsec:Dimen_Reduc_LLE}
        
        As discussed above, the physical state of a YSO can be primarily gleaned from its position in molecular column density versus excitation temperature plots. However, this description remains largely qualitative. To make it more quantitative, we can find a mapping in which the most prominent differences across YSOs manifest in their molecular excitation temperature and column density data. We can then use this embedding to classify our sources. Later, this information can help determine if spectral PCA components can classify the sources directly without the need for line hunting and rotational diagram fitting.

        
        LLE \citep{Roweis_2000} is a dimensionality reduction technique that can capture the underlying structure of high-dimensional data that lies on or near a lower-dimensional surface. LLE finds a lower-dimensional representation that preserves the local relationships between data points. The method first identifies a set of $k$-nearest neighbours for each data point in the high-dimensional space. Then, it reconstructs each data point as a weighted linear combination of its neighbours. These weights are determined so that the reconstruction minimizes the difference between the original point and its reconstructed version. 
        
        Since the relationships between data points are often locally linear, nearby points can be represented as linear combinations of each other. By preserving these local linear relationships in the lower-dimensional space, LLE creates a mapping that transforms complex non-linear patterns into a new set of coordinates, making it easier to cluster these patterns. The mapping conserves only the local correlations.

        In contrast, PCA is limited to linear transformations and may not effectively capture nonlinear patterns. The primary hyperparameter for PCA (i.e. the number of dimensions) can be easily constrained based on the desired variance to be preserved and by metrics such as the Bayesian Information Criterion (BIC) and the Akaike Information Criterion (AIC). However, the hyperparameters for LLE (number of neighbours and number of components) cannot be constrained similarly because the variance is not preserved in the projection. 
        
        When choosing hyperparameters for LLE, it is important to find a balance between capturing the local structure of the data and avoiding overfitting. For noisy data, it is recommended to start with a number of neighbours between 10 and 30. The number of dimensions is determined using two approaches. The first approach involves comparing with prior knowledge about the expected number of dimensions, which in our case is low, as indicated by PCA analysis of the spectra. The second approach involves qualitative cross-validation, where different values are experimented with, and the mapping results are visually evaluated after a downstream task such as GMM classification. Choosing appropriate hyperparameters will lead to well-behaved GMM clustering results, visually definite groups and well-defined BIC/AIC minima. LLE is also more computationally expensive than PCA. This makes it more suitable for analyzing datasets with lower dimensionality.

    \subsection{Gaussian mixture models}
    \label{subsec:Dimen_Reduc_GMM}
        Although the projection of the properties of the sources--in our case, the rotational diagram data--into the LLE-projected space may reveal structure and similarities across the sample, this projection needs to be complemented by an unsupervised classification scheme that identifies similar data clusters of YSOs. 
        Due to its versatility and ease of model selection, we use GMMs for this unsupervised classification.
        
        A GMM \citep{McLachlan_1988} is a probabilistic model used in statistics and machine learning to represent a dataset as a combination of multiple Gaussian (normal) distributions. The model estimates the parameters (means, covariances, and mixing coefficients) that best fit the data using the Expectation-Maximization (EM) algorithm. The GMM model selection can be assessed using the BIC, where GMMs are fit with different values of $K$ (the number of components) to the data. The BIC score calculation balances model fit and complexity by penalizing models with more components. The model with the lowest BIC score is selected because it represents the best trade-off between capturing the data patterns and avoiding overfitting.

    \subsection{Random forest}
    \label{subsec:Dimen_Reduc_RF}
        After the unsupervised classification step, each YSO is labelled according to non-linear similarities in their physical state data. To determine if the spectral PCA components can directly classify YSOs without line extraction and rotational diagram fitting, we use a supervised learning algorithm to create a robust and accurate predictive model.
        
        For simplicity of implementation, here we use the RF method \citep{Breiman_2001}, an ensemble learning technique used in machine learning for supervised classification tasks. It operates by constructing multiple decision trees during training, where each tree is trained on a random subset of the data (bootstrap samples) and a random subset of features. During prediction, each tree in the forest independently provides an outcome, and the final prediction is determined by a majority vote (for classification) of these individual tree predictions.
        
        When validating an RF model, techniques like cross-validation are used. This involves dividing the dataset into training and testing subsets multiple times and evaluating the model performance on each split. Here, we use metrics such as precision, accuracy, and F1-score to assess the performance of a model based on independent validation data. These metrics convey a numerical evaluation (ranging from 0-1) of the ability of the model to predict both positive and negative classes correctly. Precision reflects the accuracy of positive predictions, recall indicates the ability to avoid false positives, and the F1-score is the harmonic mean of the other two metrics. When combined, they help assess the model's effectiveness \citep{sokolova}. Scores below $1/n_\text{classes}$ are no better than a random guess, and general guidelines suggest that model metrics between 0.7 and 0.8 are appropriate in preliminary analyses. Metrics above 0.8 indicate a good performance, and the model effectively distinguishes between classes. 
        
\section{Results}
\label{sec:Results}

    We present the results of dimensionality reduction using PCA and the unsupervised classification with LLE and GMM. To validate the effectiveness of spectral PCA in representing the physicochemical properties of the sources, we use the LLE+GMM classification results with a supervised RF model. We also make predictions on a test sample (9 out of 41) that was not used in the unsupervised classification training.

    \subsection{Dimensionality reduction}
    \label{subsec:Result_Dimen_Reduc}

        In this section, we describe the results of the dimensionality reduction performed with PCA (Sect.~\ref{subsec:Dimen_Reduc_PCA}) to the 41 spectra listed in Fig.~\ref{fig:norm_spectra_sample1} and Table~\ref{tab:sources}.

        \subsubsection{Average spectrum}
        \label{subsec:Result_mean_spectrum}

            Firstly, the average spectrum of the 41 spectra (top panel of Fig.~\ref{fig:PCA_results}) is subtracted from the spectral sample to centre the data and implement the dimensionality reduction with PCA. The average spectrum represents typical features of an HMC, characterized by bright lines and rich complex molecule emissions. It contains the dominant emission of the \ch{CH3OH} $5{-}4$ ladders at the rest frequencies \mbox{${\sim}241.25$~GHz} and \mbox{${\sim}241.75$~GHz}, and the \ch{CH3CN} $13{-}12$ ladder between the rest frequencies \mbox{$238.84{-}239.13$~GHz}.

        \subsubsection{The PCA Components}
        \label{subsec:Result_PCA}
        
\begin{figure}
             \centering
                \includegraphics[width=0.98\columnwidth]{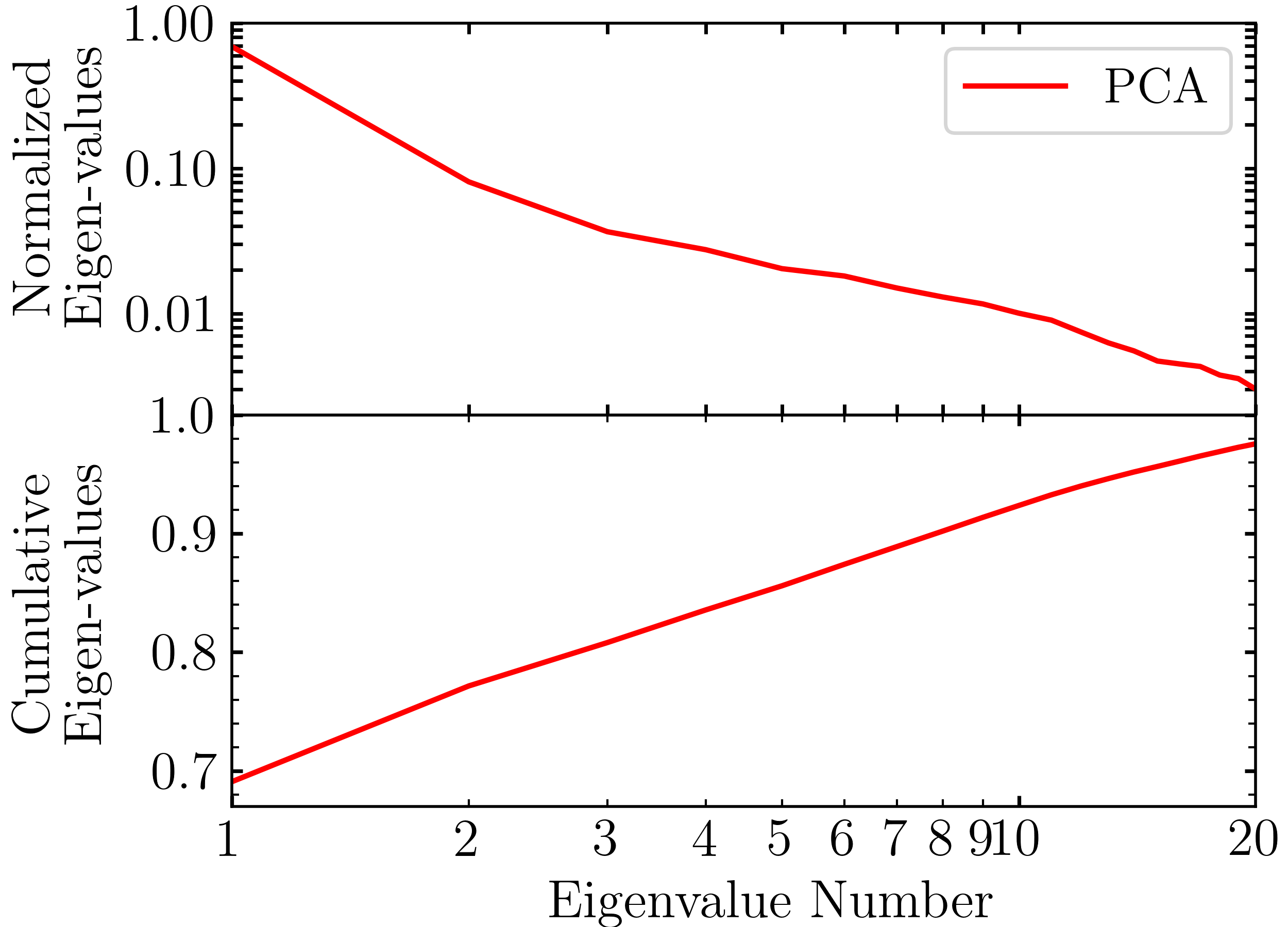}
                \caption{Normalized explained variance ratio (top panel) and cumulative explained variance ratio (bottom panel) as a function of the eigenvalue number.}
                \label{fig:PCA_frac_var}
            \end{figure}

\begin{figure*}
             \centering
                \includegraphics[width=\textwidth]{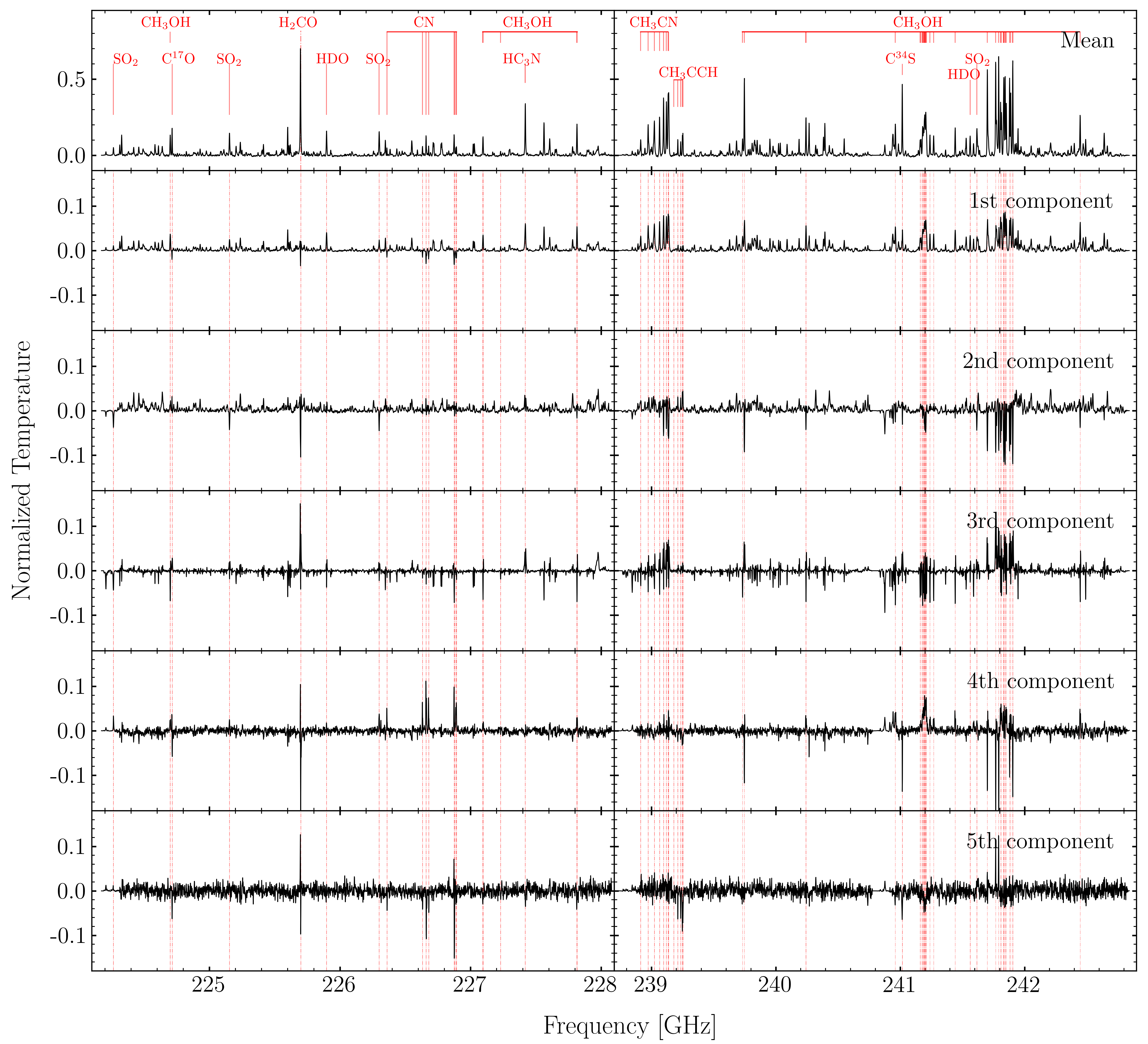}
                \caption{The top panel is the mean spectrum of the 41 sources selected for PCA (Table~\ref{tab:sources}). The other panels are the first five principal components of the PCA. Some molecular lines are marked at their central LSR frequency according to the CDMS catalogue.}
                \label{fig:PCA_results}
            \end{figure*}

            We performed the dimensionality reduction with 41 spectra (Fig.~\ref{fig:norm_spectra_sample1}) using PCA\footnote{\url{https://scikit-learn.org/stable/modules/generated/sklearn.decomposition.PCA.html}} function of the \texttt{scikit-learn} library \citep{Pedregosa_scikit-learn_2011}. PCA successfully reduced the high-dimension problem--at least--to eight new dimensions (i.e.~eigenspectra), explaining approximately 90\% of the total variance. The partial sum of the explained variance is presented in Fig.~\ref{fig:PCA_frac_var} and Table~\ref{tab:PCA_CumSum_values}. The resulting eigenspectra are shown in Fig.~\ref{fig:PCA_results}.             

            The first eigenspectrum (Fig.~\ref{fig:PCA_results}, second row) represents ${\sim}$69\% of the total sample variance and has a high degree of similarity to the average spectrum of the sample. However, the \ch{CH3OH} $5{-}4$ ladders at \mbox{${\sim}241.25$~GHz} and \mbox{${\sim}241.75$~GHz}, and the \ch{CH3CN} $13{-}12$ ladder between \mbox{$238.84{-}239.13$~GHz} seem broader and in correspondence with optically thick emission and high temperatures. The described characteristics are more evident within the \ch{CH3OH} $5{-}4$ ladders. Therefore, the first principal component may represent bright, dense, hot sources (i.e.~HMCs); see the discussion in Sect.~\ref{subsec:Discu_phy_param_compar} for more details.

            %
            \begin{table}
             \caption{Partial sums of the weights of the sample's eigenspectra.}
             \label{tab:PCA_CumSum_values}
             \centering
             \resizebox{\columnwidth}{!}{
             \begin{tabular}{c c c c c c}
              \hline\hline
              Comp. & Var. \% & Cum. sum \% & Comp. & Var. \% & Cum. sum \% \\
              \hline
              1 & 69.09 & 69.09 & 11 & 0.90 & 93.26 \\
              2 & 8.07 & 77.16 & 12 & 0.74 & 94.01 \\
              3 & 3.66 & 80.81 & 13 & 0.63 & 94.63 \\
              4 & 2.75 & 83.56 & 14 & 0.55 & 95.18 \\
              5 & 2.03 & 85.59 & 15 & 0.47 & 95.65 \\
              6 & 1.81 & 87.40 & 16 & 0.45 & 96.10 \\
              7 & 1.50 & 88.90 & 17 & 0.43 & 96.53 \\
              8 & 1.30 & 90.20 & 18 & 0.38 & 96.91 \\
              9 & 1.16 & 91.36 & 19 & 0.36 & 97.26 \\
              10 & 1.00 & 92.36 & 20 & 0.30 & 97.56 \\
              \hline
             \end{tabular}
             }
            \end{table}
            %
            %
            \begin{figure}
             \centering
                \includegraphics[width=\columnwidth]{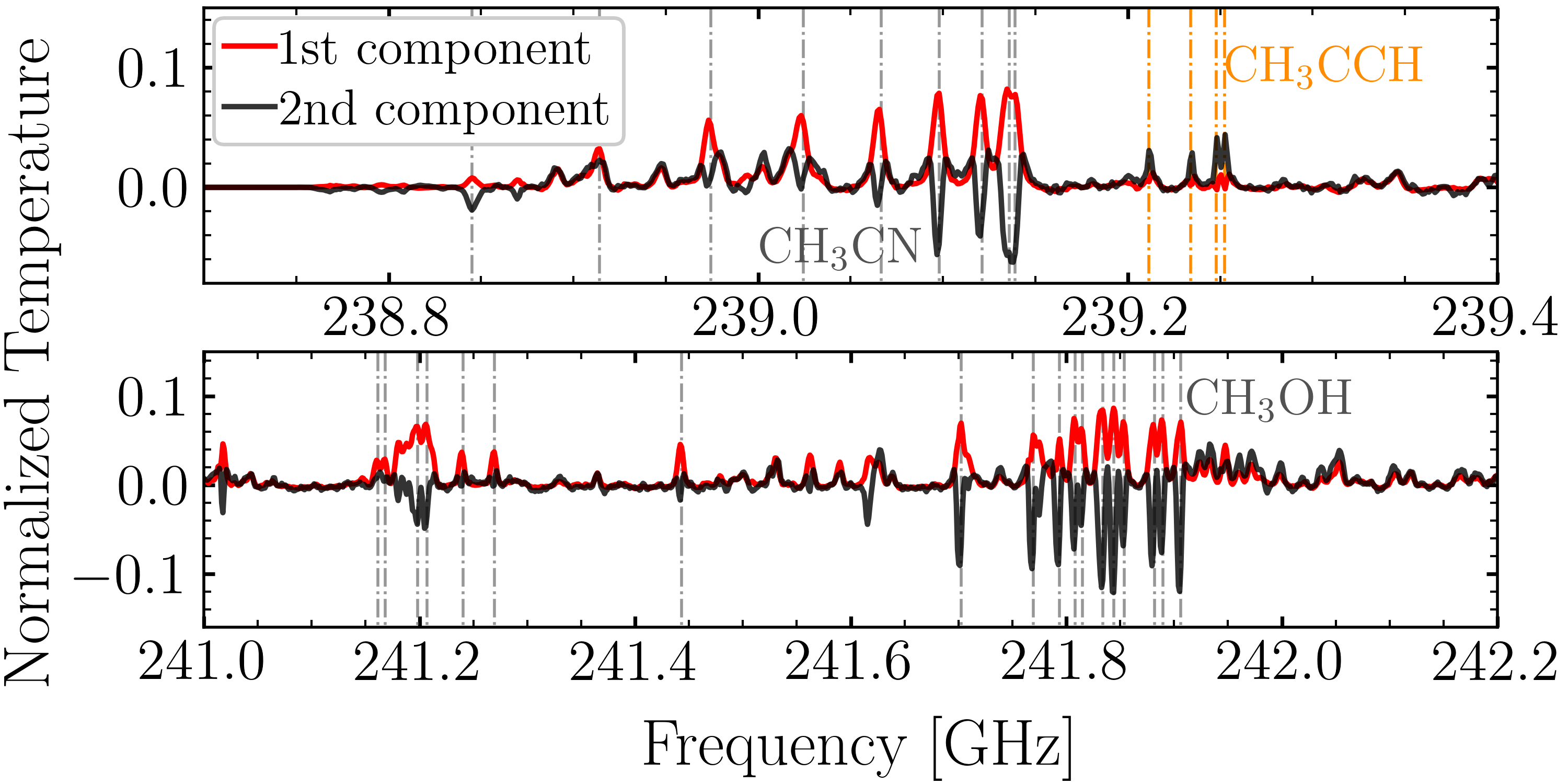}
                \caption{Comparison between the first (red line) and second (black line) eigenspectra at the frequencies of the \ch{CH3CN} $13{-}12$ ladder and \ch{CH3CCH} lines (top), and the \ch{CH3OH} $5{-}4$ ladders (bottom).}
                \label{fig:1st-2nd_comp_SO2-CH3CN}
            \end{figure}

            The second eigenspectrum (Fig.~\ref{fig:PCA_results}, third row) represents ${\sim}$8\% of the total sample variance (see also Table~\ref{tab:PCA_CumSum_values}). Its general characteristic is the turnover of some bright lines such as the \ch{CH3OH} and \ch{CH3CN} ladders, which are shown in detail in Fig.~\ref{fig:1st-2nd_comp_SO2-CH3CN} (in black) compared with the first eigenspectrum (in red). The lines' turnover between eigenspectra of different orders does not necessarily represent absorption features. The eigenspectra are orthogonal by definition. Therefore, both positive and negative contributions are expected. These turnovers could either be a connection between the PCA components and the presence or absence of these molecules within the MYSOs or an issue with the alignment along the frequency axis of the spectral sample; see discussion in Sect.~\ref{subsec:Discu_pseudo_spec_interp}.
            
            Furthermore, the emission of the \ch{CH3CCH} $14{-}13$ ladder between the rest frequencies \mbox{$239.13{-}239.25$~GHz} is enhanced in the second PCA component (Fig.~\ref{fig:1st-2nd_comp_SO2-CH3CN}, in orange). As mentioned in Sect.~\ref{subsec:COMs_selected}, \ch{CH3CCH} is usually found in the envelopes of MYSOs, and that is the case of the sample studied here \citep[see][for more details]{Frimpong_2021}. Since the bright \ch{CH3OH} and \ch{CH3CN} features are better represented by the first component and it is opposite to the second component (i.e.~there is a turnover), the second eigenspectrum may be associated with the emission of the cold surroundings of the MYSOs.
            
            Finally, the third through eighth eigenspectra account for less than $3.7\%$ of the variance each (Table~\ref{tab:PCA_CumSum_values}). No clear pattern is observed in these components regarding the COMs studied here, except for a turnover of the \ch{CH3CCH} $14{-}13$ ladder and the enhancement of the \ch{CH3OH} $5{-}4$ ladder in the fourth component (see the fifth row in Fig.~\ref{fig:PCA_results}). It is also possible that high-order components may be tracing properties associated with molecules different from those studied here. Nevertheless, the higher the order, the more noisy the PCA component becomes, indicating that high-order eigenspectra may also represent noise features.

    \subsection{Classification}
    \label{subsec:Result_classif}

        \subsubsection{LLE and GMM}
        \label{subsec:Result_LLE_GMM}
            We found an LLE projection of the 8-dimensional molecular excitation and physical dataset for the gas mass, total column density,  \ch{CH3OH}, \ch{CH3CN}, \ch{CH3CCH} rotational temperature and column density for the 32 YSOs in our sample for which all three rotational diagrams could be constructed (Table~\ref{tab:COMs_param}). We used a MinMax scaler on the data and ran the \texttt{scikit-learn} LLE algorithm with $n=20$ neighbouring points, reducing the dimensionality of the data from $8$ to $2$. The choice of the number of dimensions is supported by the first two PCA components covering almost 80\% of the variance and by qualitative cross-validation of the GMM clustering results. The projected data shows a correlation with COM temperature, which means that the presence or absence of COMs explains most of the LLE-projected variance.
    
\begin{figure}
             \centering
                \includegraphics[width=0.98\columnwidth]{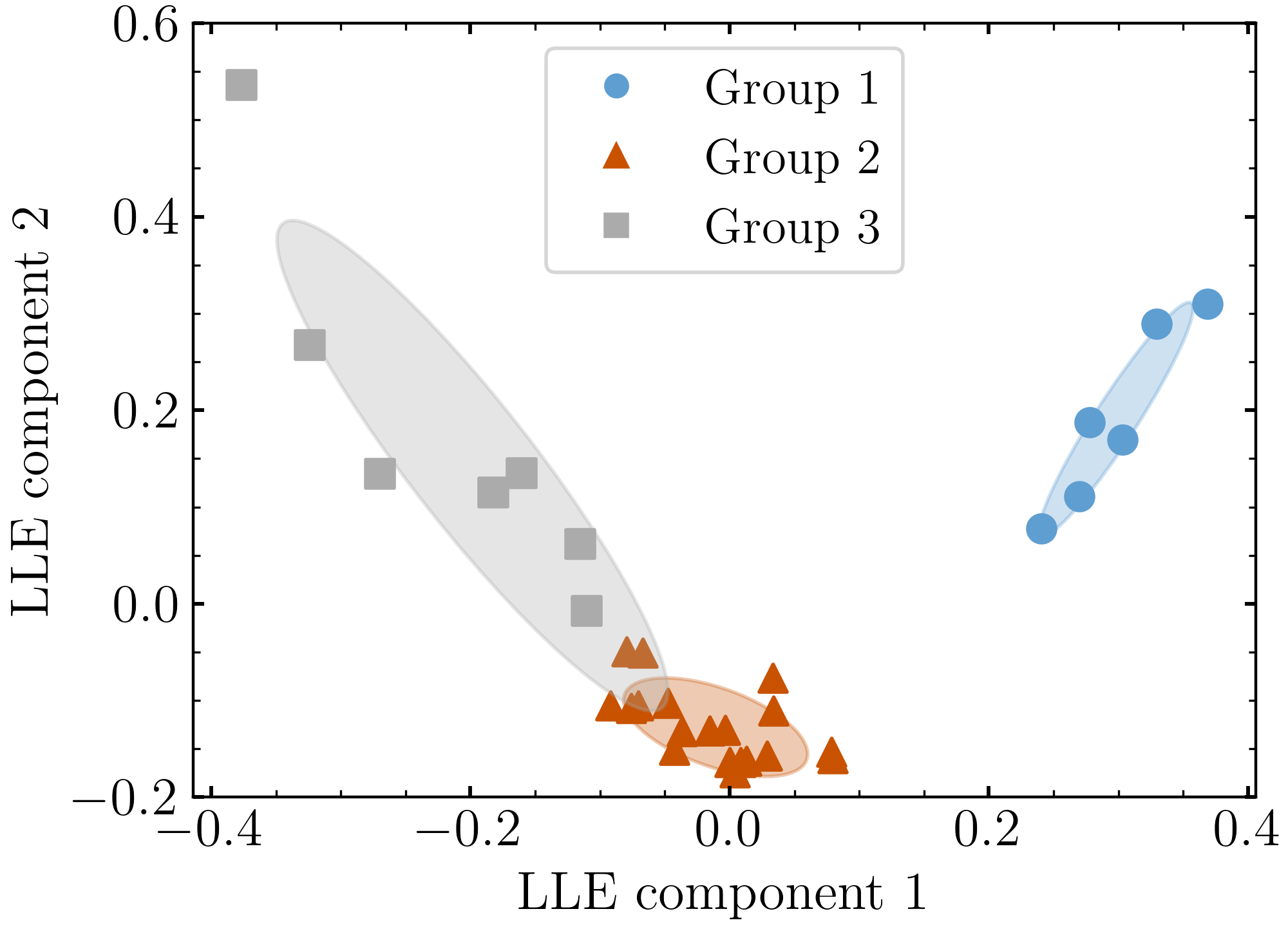}
                \caption{Two-component LLE projection of gas density, mass, and \ch{CH3OH}, \ch{CH3CN}, \ch{CH3CCH} rotational temperature and column density for 32 YSOs for which the rotational diagram yielded this information. The colours correspond to the labels assigned by a three-component, BIC-optimized GMM clustering. The ellipses encompass 95.4\% ($2\sigma$) of the probability mass of each GMM cluster.}
                \label{fig:LLEGMM}
            \end{figure}

            We then looked for similarities between clusters of LLE-projected data points corresponding to similar YSOs. We used GMM and selected the model using the BIC from the \texttt{scikit-learn} library. We thus found that a 3-cluster, full covariance model minimizes the BIC. The results of the dimensionality reduction plus clustering are shown in Fig.~\ref{fig:LLEGMM}.
    
            The GMM clustering results show that the three assigned labels adequately divide the sources in LLE space, although there is some overlap between classes. Table~\ref{tab:LLEGMM} summarises the physical properties of each of these classes. Group~1 corresponds to cold ($T\lesssim40$~K) and COM-poor YSOs, Group~2 to warm ($T{\sim}200{-}350$~K), diffuse \mbox{($N_\text{gas}{\sim}1.1{-}3.5\times10^{24}$~cm$^{-2}$)}, low-mass, and medium-COM-abundance YSOs, and Group~3 to warm-to-hot ($T{\sim}250{-}500$~K) and COM-rich YSOs. Figure~\ref{fig:LLEGMM} shows that an evolutionary picture begins to emerge: Group~1 sources (cold, COM-poor) could be evolving into Group~2 (warm, medium-COM-abundance), which could then evolve into Group~3 (warm-to-hot, COM-rich). 
            The $32$ sources are individually classified as shown in Table~\ref{tab:COMs_param}. These sources have been labelled with a prediction probability of (${>}93\%$) except for Group~2 sources \mbox{G339.6221-00.1209} ($72\%$) and \mbox{G345.5043+00.3480} ($66\%$). We may also surmise that Group~2 sources are well-concentrated, although around 10\% of them may be wrongly assigned to Group~3. Also, while Group~1 sources are better isolated than those in Group~3, sources identified as belonging to these groups are almost unequivocally identified as such (prediction probabilities approximating 1).            
            \begin{table*}
            \caption{\label{tab:LLEGMM}Results from the LLE+GMM unsupervised classification.}
            \centering
            \begin{tabular}{c c c c c c c c c c}
                \hline\hline
                (1) & (2) & (3) & (4) & (5) & (6) & (7) & (8) & (9) & (10) \\
               GMM  & $T$  & $M_\text{gas}$  & $N_\text{gas}$  & $T_\text{CH3OH}$  & $N_\text{CH3OH}$   & $T_\text{CH3CCH}$  & $N_\text{CH3CCH}$  & $T_\text{CH3CN}$ & $N_\text{CH3CN}$ \\ 
               group& (K) & ($M_\odot$) & ($10^{24}$~cm$^{-2}$) &  (K) & ($10^{16}$~cm$^{-2}$) & (K) & ($10^{16}$~cm$^{-2}$) & (K) & ($10^{16}$~cm$^{-2}$) \\
               \hline               
                1&
                $20_{-6}^{+18}$&
                $34_{-6}^{+12}$&
                $8.4_{-3.4}^{+12.6}$&
                $20_{-6}^{+10}$&
                $0.13_{-0.06}^{+0.34}$&
                $30_{-16}^{+2}$&
                $0.55_{-0.31}^{+0.73}$&
                $30_{-16}^{+16}$&
                $0.004_{-0.002}^{+0.033}$
                \\ [0.3cm] 
                2&
                $220_{-20}^{+130}$&
                $7.8_{-3.3}^{+10.2}$&
                $1.9_{-0.8}^{+1.7}$&
                $220_{-34}^{+34}$&
                $25_{-19}^{+76}$&
                $60_{-11}^{+20}$&
                $0.65_{-0.31}^{+0.36}$&
                $220_{-20}^{+130}$&
                $0.5_{-0.4}^{+1.5}$ 
                \\ [0.3cm] 
                3&
                $300_{-40}^{+200}$&
                $26_{-14}^{+18}$&
                $5.6_{-3.2}^{+1.4}$&
                $250_{-20}^{+4}$&
                $300_{-51}^{+220}$&
                $75_{-25}^{+15}$&
                $1.6_{-0.8}^{+0.7}$&
                $300_{-40}^{+200}$&
                $6.6_{-3.6}^{+3.4}$
                \\ \hline
            \end{tabular}
            \tablefoot{Columns: (1) Unsupervised classification group identification; (2), (3), and (4) Gas temperature, mass, and column density, respectively; rotational temperatures and column densities of \ch{CH3OH} (5) and (6), \ch{CH3CN} (7) and (8), and \ch{CH3CCH} (9) and (10), respectively, for the 32 YSO subsample for which rotational diagrams could be constructed.
            }
            \end{table*}

        \subsubsection{PCA validation with RF}
        \label{subsec:Result_PCA_val}
            To test the usefulness of using spectral PCA for molecular analysis, we used the labels from the LLE+GMM classification of YSOs to train an RF model using the first five PCA components of the same 32 sources used in the previous analysis. We used the following hyperparameters for the RF model following the literature recommendations in \cite{rf_book} 20 jobs, 100 estimators, 1/3 test/train split. We also evaluate the robustness of the model with a 10-fold cross-validation. This simple RF model can accurately predict the LLE+GMM labels obtained via the manual line extraction and rotational diagram data from the spectral PCA components. The resulting confusion matrix of the model, with 1000 bootstrap realizations for getting uncertainties is,
            \begin{equation*}
            \left[
            \begin{array}{c|ccc}
             & \text{Pred 1} & \text{Pred 2} & \text{Pred 3} \\
            \hline
            \text{Actual 1} & 8.87\pm 2.24 & 0 & 2.96\pm1.59 \\
            \text{Actual 2} & 0 & 4.05\pm1.80 & 0 \\
            \text{Actual 3} & 2.07\pm1.38 & 0 & 2.04\pm1.37 \\
            \end{array}
            \right]\ .
            \end{equation*}
            This corresponds to the following weighted average metrics: \mbox{Precision $= 0.77$}, Recall $= 0.75$, and F1-score $= 0.76$ with the PCA components one and two contributing ${\gtrsim}50$\% of feature importance for the RF model. Similarly to the GMM classification above, Class 2 is well isolated, but Classes 1 and 3 are mixed, which means that the RF model has lost some information. We emphasize that the simple RF model used here is presented as a proof of concept that PCA can help directly classify sources at a level comparable with manual, individual line extraction. The performance of the RF model shown here is indicative that deep learning models may be able to accomplish this task with less confusion. However, finding such a model is beyond the scope of this work.

        \subsubsection{PCA predictions}
        \label{subsec:Result_PCA_predic}
            We can now use the RF model to classify the previously unclassified sources. Among the 41 sources, nine lacked excitation temperature and column density data. Therefore, we used their first five PCA eigenvectors to predict their classification. The RF model provided supervised classifications, as detailed in Table~\ref{tab:COMs_param}, with the prediction probability indicated in parentheses.

\section{Discussion}
\label{sec:Discu}

    \subsection{Eigenspectra interpretation}
    \label{subsec:Discu_pseudo_spec_interp}
        \ch{CN} ``absorption'' features between the rest frequencies \mbox{$226.2{-}227.0$~GHz} are quite evident in the first principal component (Fig.~\ref{fig:PCA_results}). The same features are observed in the real spectra of sources such as \mbox{G333.0682-00.4461}, \mbox{G332.9636-00.6800}, \mbox{SDC22.985-0.412\_1}, \mbox{SDC23.21-0.371\_1}, \mbox{SDC33.107-0.065\_2}, and \mbox{SDC43.311-0.21\_1}. Most are bright objects (e.g.~see ~Fig.~\ref{fig:norm_spectra_sample1}). However, such features may have multiple reasons: i) an artefact of interferometric observations that may appear in the process of filtering the continuum, which can happen when the object has abundant extended material around; (ii) cold diffuse \ch{CN} from the ISM between the source and the observer, thus, a difference in the velocity of the line should be observed, but that is not the case; or (iii) cold structures or clumps that belong to the same molecular cloud where the star formation is happening, but are independent of the central hot region and are settled in the line of sight, this could also explain the shared velocity with the central source. 

        Nevertheless, care must be taken in interpreting apparent absorption or emission in the eigenspectra since the turnover of a molecular line across the eigenspectra of different orders may or may not mean a real feature in an individual source. As explained above, eigenspectra are orthogonal by definition, meaning that they can have positive and negative values. There may be cases when an apparent absorption counterpart to a higher-order eigenspectrum emission feature appears (e.g.~\ch{CN} lines). The same applies to apparent emission counterparts to higher-order eigenspectrum absorption features (e.g.~\ch{CH3OH} or \ch{CH3CN}). However, molecular line features in eigenspectra do not inform us about actual emission/absorption but merely of the existence of (and possible correlations among) molecules associated or not with the eigenspectra.

        Furthermore, the turnover in some features across eigenspectra of different orders might be due to frequency alignment issues in the observed spectra. For example, we applied a shift using a single $V_\text{lsr}$ per source (Sect.~\ref{subsec:post-reduc_LSR_dopplerShift}), but some species may have different velocities or multiple velocity components. Additionally, some species may present self-absorption, adding an extra complication to the spectra alignment.
        %
        \begin{figure*}
         \centering
             \begin{subfigure}[b]{\textwidth}
                 \centering
                 \includegraphics[width=\linewidth]{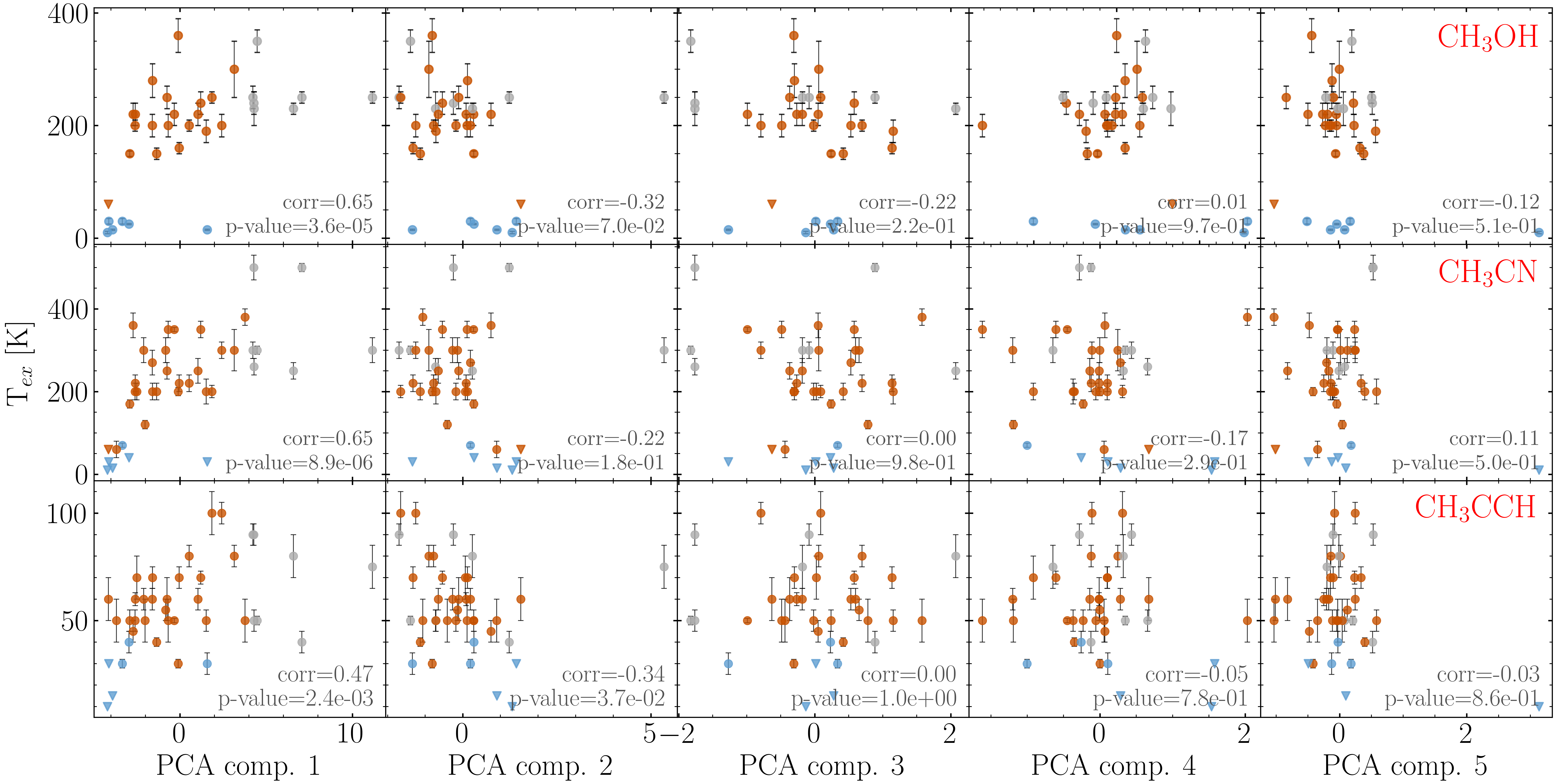}
             \end{subfigure}
             \hfill
             \begin{subfigure}[b]{\textwidth}
                 \centering
                 \includegraphics[width=\linewidth]{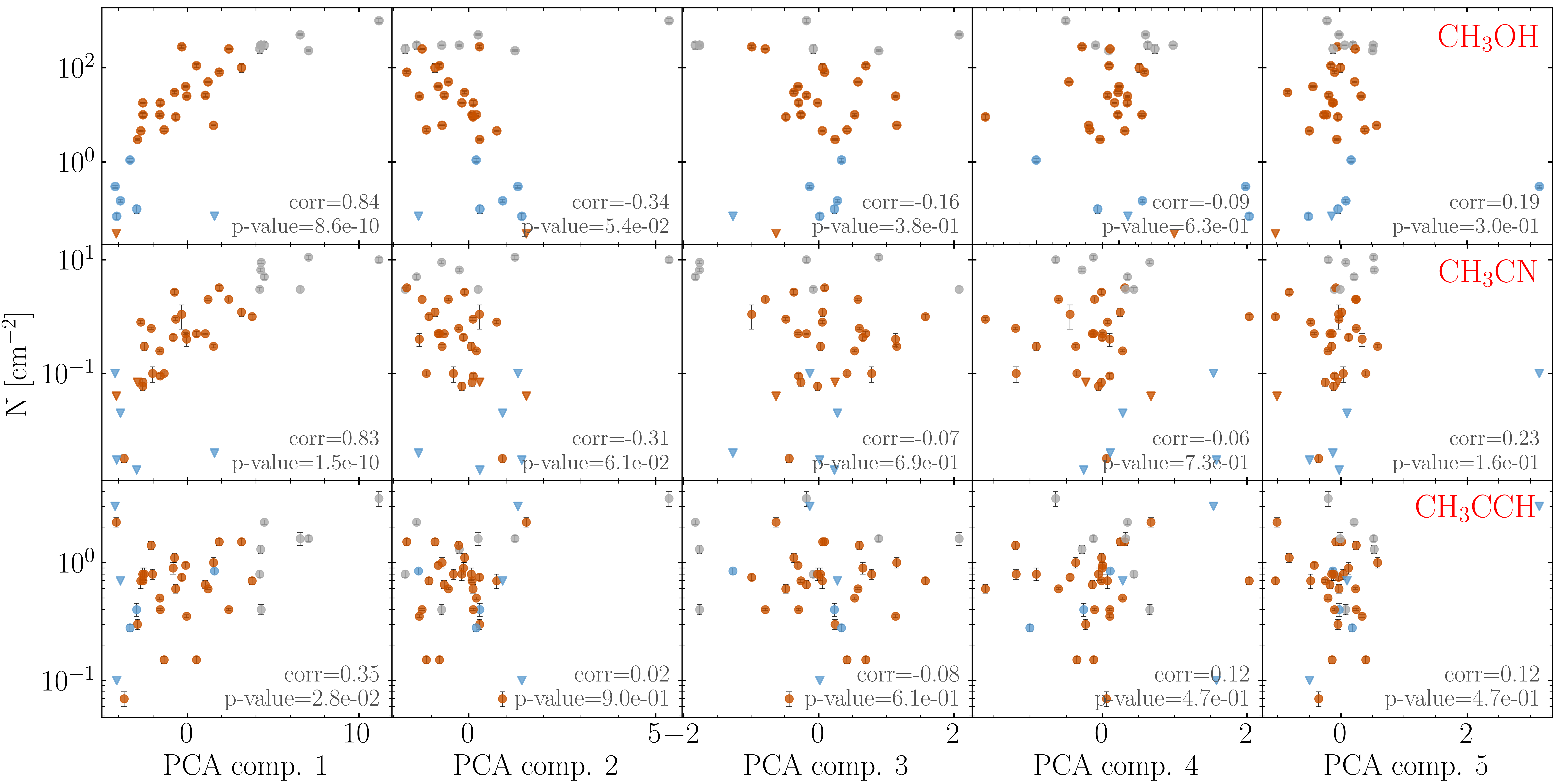}
             \end{subfigure}
             \caption{Excitation temperatures (top panels) and column densities (bottom panels) of \ch{CH3OH} (top row), \ch{CH3CN} (middle row), and \ch{CH3CCH} (bottom row) as a function of the first five PCA eigenvector components. The values are coloured by the groups' classification found in Sect.~\ref{subsec:Result_classif}: Group~1 (the blue dots), Group~2 (the brown dots), and Group~3 (the grey dots).}
             \label{fig:pca_vs_Tex_N}
        \end{figure*}

    \subsection{Correlations with physical and chemical parameters}
    \label{subsec:Discu_phy_param_compar}

        We compared the PCA eigenvector components with the excitation temperature and column density of \ch{CH3OH}, \ch{CH3CN}, and \ch{CH3CCH} (Fig.~\ref{fig:pca_vs_Tex_N}), and found patterns consistent with the correlations observed by \cite{Frimpong_2021} and \cite{Frimpong_2023subm}. The first PCA component (Fig.~\ref{fig:pca_vs_Tex_N}, left column) shows a significant correlation with the excitation temperatures and column densities of \ch{CH3OH} (top row) and \ch{CH3CN} (middle row); see correlation coefficients and p-values on each panel of Fig.~\ref{fig:pca_vs_Tex_N}. On the other hand, the \ch{CH3CCH} excitation temperature and column density (bottom row) display a marginal ${<}3\sigma$ correlation with the first PCA component. The clear correlations of $T_{\ch{CH3OH}}$, $T_{\ch{CH3CN}}$, $N_{\ch{CH3OH}}$, and $N_{\ch{CH3CN}}$ with the first component confirm that the hot emission of the COMs dominates the first eigenspectra.

        Furthermore, the second PCA component shows a marginal anti-correlation with the temperature and column density of the COMs studied. Most sources where hot emission was detected ($T{>}100$~K) contribute negatively to the second PCA component, whose eigenspectrum profile is dominated by colder MYSOs ($T{<}100$~K). Finally, higher-order components do not show any particular trend related to the excitation temperature or column density of the studied COMs. 

        Figure~\ref{fig:pca_vs_Tex_N} includes the classification found with LLE and GMM (Sect.~\ref{subsec:Result_classif}), using the same colour code as in Fig.~\ref{fig:LLEGMM}. The groups are clearly separated along the correlations of Fig.~\ref{fig:pca_vs_Tex_N} in the first and--marginally--in the second PCA component. Let us recall that the LLE and GMM unsupervised classification was made with the physical and chemical properties known in a subsample (Table~\ref{tab:COMs_param}), that is, the classification is not directly related to the spectral profiles as the PCA results. Therefore, it is independent of the PCA components. 
        %
        \begin{figure*}
         \centering
                \includegraphics[width=\linewidth]{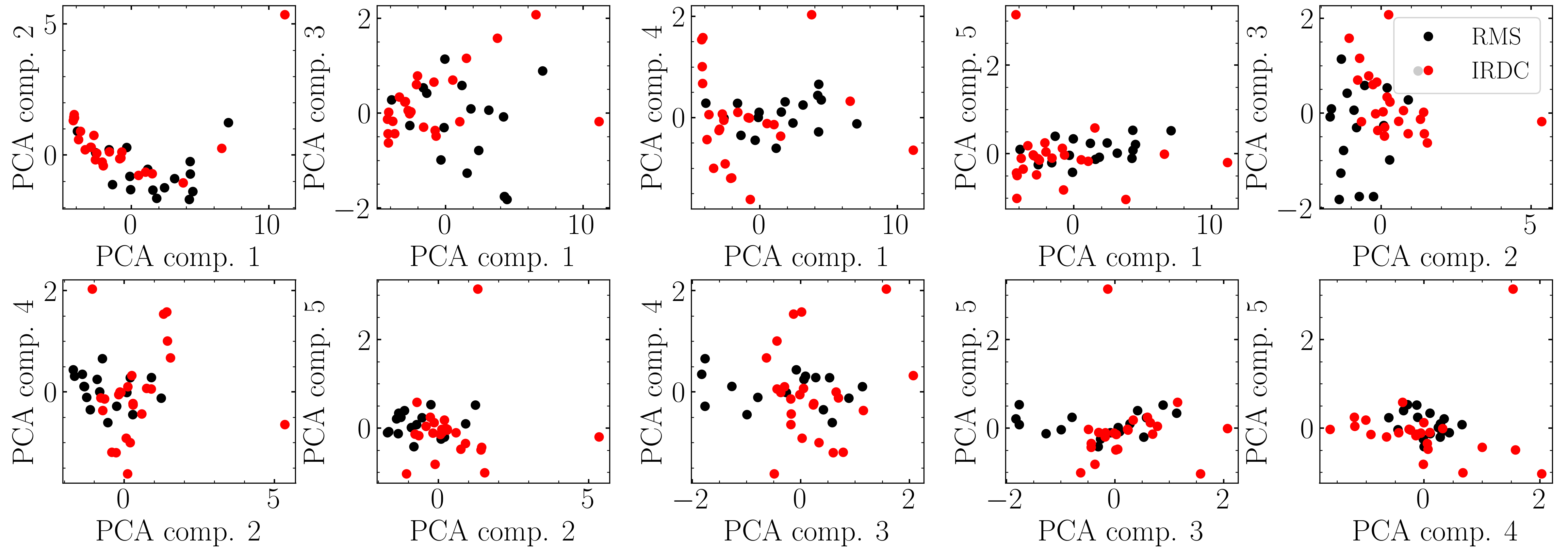}
             \caption{PCA eigenvector components comparison. RMS sources are in red, and IRDC sources are in black.}
             \label{fig:PCA_coeff_compar}
        \end{figure*}
        
        Group~3 (the grey dots), corresponding to sources with the highest COM temperatures and column densities, stands out as the only group with a wide range of coefficient values (both negative and positive) in the second PCA component. This group often has approximately the same temperature and column densities of \ch{CH3OH} and \ch{CH3CN}, suggesting that Group~3 is not well represented by the second PCA component.

    \subsection{Chemical evolution patterns}
    \label{subsec:Discu_chemi}

        Figure~\ref{fig:PCA_coeff_compar} shows the distributions between the first five PCA eigenvector components coloured by the catalogues (Sect.~\ref{sec:obs_data}). The catalogue identification (i.e.~RMS and IRDCs) does not have a clear pattern in the distribution of most of the panels of Fig.~\ref{fig:PCA_coeff_compar}. However, the comparison between the first and the second PCA components (the top-left panel) shows the eigenvectors of the IRDC sources preferably clustered to the left and those of the RMS sources to the right of the panel. The pattern found between the first two PCA components thus may indicate an evolutionary difference between the objects of the two catalogues, as expected. Nevertheless, both catalogues might have--at some level--a mixture of MYSOs in different evolutionary stages. 
        %
        \begin{figure}
         \centering
            \includegraphics[width=0.98\columnwidth]{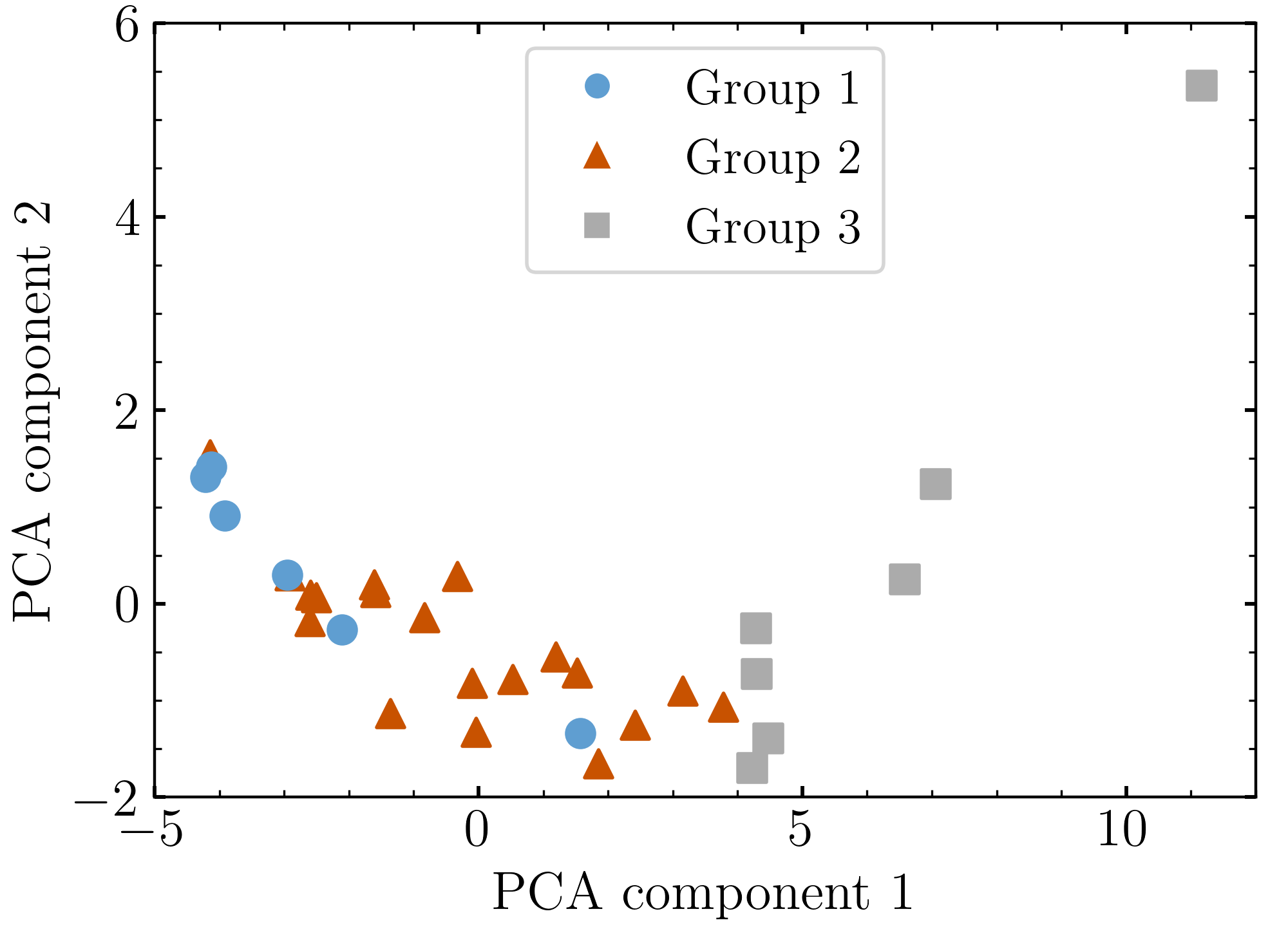}
            \caption{Second PCA component as a function of the first PCA component coloured by the group classification found in Sect.~\ref{subsec:Result_classif}.}
            \label{fig:pc1_vs_pc2_3groups}
        \end{figure}
        %
        \begin{figure*}
         \centering
            \includegraphics[width=\textwidth]{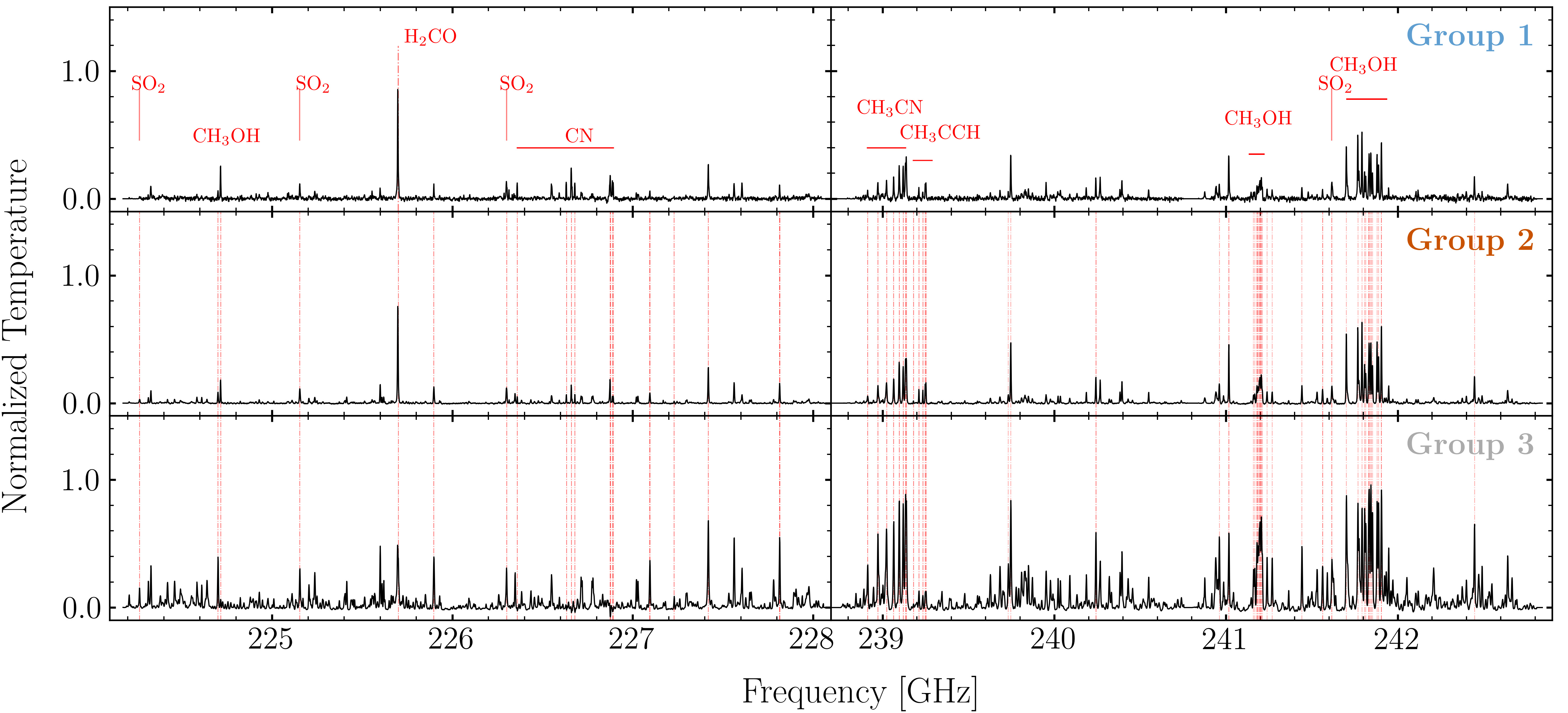}
            \caption{Average spectra of the three groups of sources found in Sect.~\ref{subsec:Result_classif}: Group~1 are Cold and COM-poor sources, Group~2 are warm and medium-COM-abundance sources, and Group~3 are warm-to-hot and COM-rich sources.}
            \label{fig:averg_spec_3groups}
        \end{figure*}

        Considering the group classification obtained with the LLE+GMM method (Sect.~\ref{subsec:Result_classif}), we see again a pattern described by the first and second PCA components in Fig.~\ref{fig:pc1_vs_pc2_3groups}. The three groups distinguished by the unsupervised classification techniques lie along the distribution in Fig.~\ref{fig:pc1_vs_pc2_3groups}, meaning that the first two PCA components, which rely only on the spectra information, are closely related to the chemical evolution of the sample. We computed the average spectrum of each group (Fig.~\ref{fig:averg_spec_3groups}) and found that the chemical richness and line brightness increased gradually from Group~1 to Group~2 and then Group~3. 
        
        The groups' distribution in Fig.~\ref{fig:pc1_vs_pc2_3groups} is a clear signal of chemical evolution from a ``weak line'' group consisting of cold and optically thick sources to a ``bright line'' group of hotter and less embedded objects with spectra similar to HMCs. The average spectrum of Group~2 lies in an intermediate evolutionary stage compared to the other two groups and represents a mixture of RMS and IRDC objects. Our conclusions agree with previous studies \citep{Frimpong_2021}, which have suggested similar evolution patterns, identifying two distinct stages with hints of a third, intermediate stage. Other studies have used different dimensionality reduction techniques with larger spectral samples of MYSOs, both simulated and observed, and reached similar conclusions in the classification of spectra based on the presence or absence of emission lines \citep{Ward&Lumsden_2016}. It is noteworthy that these conclusions also became evident in our small sample thanks to the techniques used here.

    \subsection{Caveats}
    \label{subsec:Discu_caveats}
    
        The spectral sample has few line-poor and cold sources, and some objects lack information on their physical and chemical parameters. So, the actual conditions or properties of a particular group of MYSOs may be underrepresented in the sample. Consequently, the classification is less accurate, and the prediction loses precision. A larger and more diverse spectral sample would solve the issue.
        
        A handful of sources used in the PCA prediction are two-component temperature systems (Table~\ref{tab:COMs_param}), except for object SDC35.063-0.726\_1, which is a two-component velocity system \citep{Frimpong_2021}. Since the temperature and column density rule the classification, the hottest component will dominate over the coldest in the prediction outcome if both contributions are not previously isolated. In other words, the final prediction will miss information from the cold components. One way to tackle this issue is to separate the properties of the two components before the analysis.

\section{Summary}
\label{sec:Summary}

Our low-dimensional embedding of temperature, density, source mass, molecular excitation temperature, and column density data using LLE and GMM for unsupervised classification yielded three distinct groups, pointing to different evolutionary and physical stages for 32 (of 41) sources for which molecular information was available from rotational diagram fitting. The identified groups were: 1) cold, COM-poor sources; 2) warm, diffuse, low-mass, medium-COM-abundance sources; and 3) warm-to-hot, COM-rich sources.

We show that source spectra can be used for direct spectral source classification at a level comparable to information based on manual line extraction and rotational diagram fitting yields. We achieved this by using the main eigenspectra PCA components (i.e.~covering >90\% of variance) to train a simple RF model, achieving accuracy metrics nearing 80\%. This model allowed for a classification of 9 additional sources that did not have rotational diagram-fitting results and, thus, were previously unclassifiable. From this, we surmise that the prediction and model accuracy metrics could be improved by using more sophisticated models based on deep learning techniques. We should note that although the machine learning algorithms used here may be more computationally intensive than traditional methods, our approach requires far less manual data preparation, making it more efficient overall.

After dimensionally reducing both the molecular parameter data (LLE) and the spectra (PCA), a clearer picture of evolutionary stages for MYSOs emerges: Group~1 sources (Cold, COM-poor) are likely evolving into Group~2 sources (warm, medium-COM-abundance), which may then evolve to Group~2 (warm-to-hot, COM-rich) sources. This agrees with previous studies that have hinted at a similar evolutionary picture for MYSOs.

The sample used limits our study to only two evolutionary stages of MYSOs (HMCs and IRDCs), and the cold sources are underrepresented. A more diverse sample, including, for instance, more cold clumps and \mbox{H\,{\sc ii}} regions, will significantly improve the accuracy of the unsupervised training set classification and, most importantly, the PCA outcome. Furthermore, an extended analysis of other molecular species, for example, simple and complex molecules linked to the chemical evolutionary paths of the COMs studied, will shed light on the patterns unidentified or overlooked in the PCA components.

Dimensionality reduction techniques facilitate the classification, analysis, and prediction of data from observations with high sensitivity, spatial, and spectral resolution. Our approach could also be useful for analyzing other spectral data with molecular line emission and different wavelength regimes. Additionally, the technique could be applied to various types of sources. For instance, it could help us to study the different evolutionary stages of low-mass YSOs, which can exhibit more complexity due to the emergence of shocks and outflows. We could also consider applying this method to study protoplanetary discs, although the large parameter space may be challenging.

\begin{acknowledgements}
    The authors acknowledge funding from the Science and Technology Facilities Council (STFC) through the Radio Astronomy for the Development of the Americas (RADA: Big Data Colombia) project, grant number ST/R001944/1.
    Y. A. acknowledges the support from Prof. Marijke Haverkorn and funding from the European Research Council (ERC) under the European Union’s Horizon 2020 research and innovation program (grant agreement No 772663).
\end{acknowledgements}

%
\bibliographystyle{aa} 
\bibliography{Main_aanda.bbl} 

\begin{appendix} 
\section{Complete standardized spectral sample} \label{appendix:sample}
\begin{figure*}
     \centering
        \includegraphics[width=\linewidth]{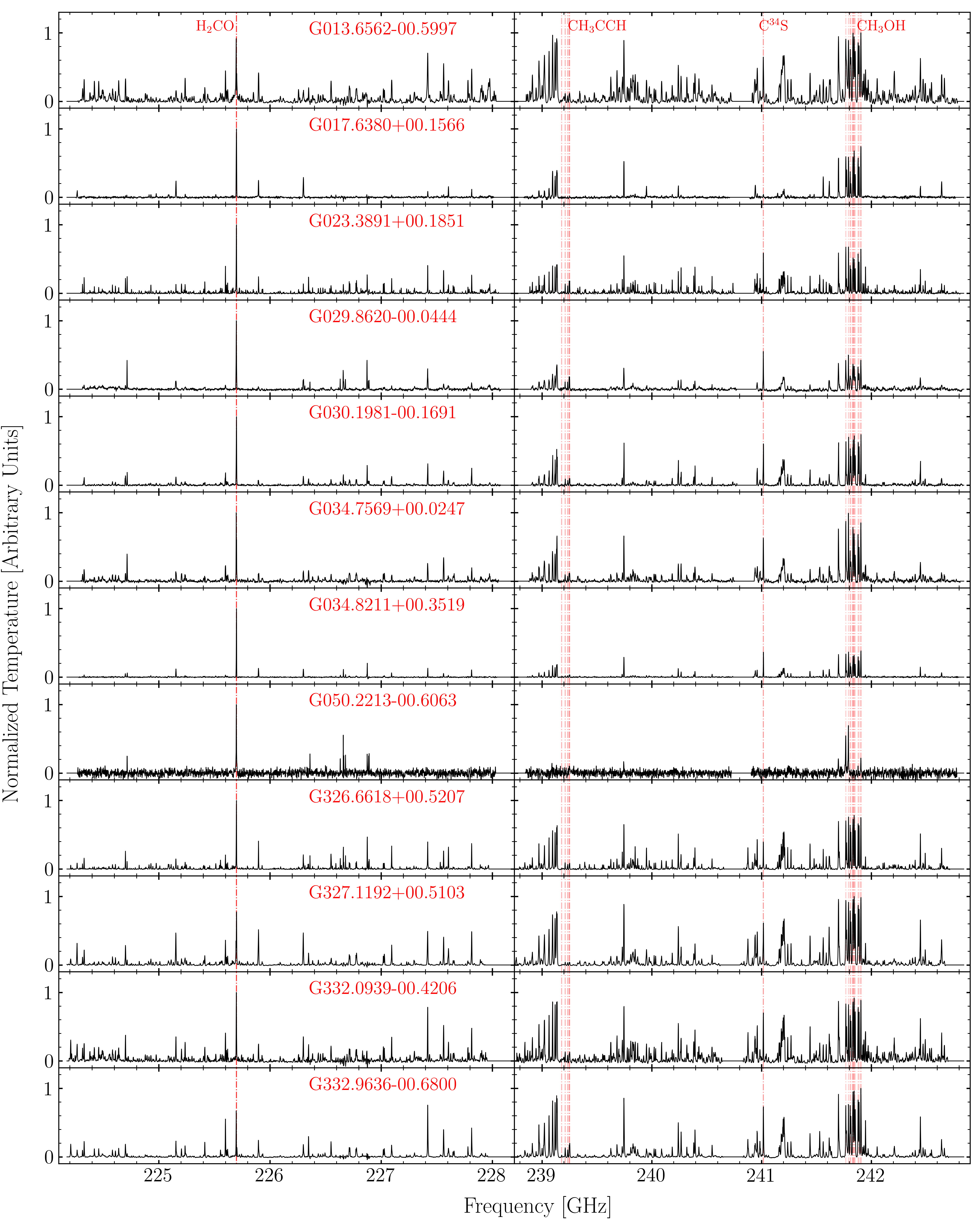}
        \caption{Same as Fig.~\ref{fig:norm_spectra_sample1}, but showing the complete normalized spectral sample of 41 MYSOs. The vertical red lines indicate the emission lines used in the normalization (See Sect.~\ref{subsec:post-reduc_norm} for more details).}
        \label{fig:norm_spectra_sample2}
    \end{figure*}

    The normalized spectral sample was presented in Sect.~\ref{subsec:post-reduc}, with five spectra displayed in Fig.~\ref{fig:norm_spectra_sample1}. Figure~\ref{fig:norm_spectra_sample2} provides the remaining normalized spectra for all 41 MYSOs.
    %
    %
    \begin{figure*}
     \centering
             \includegraphics[width=\linewidth]{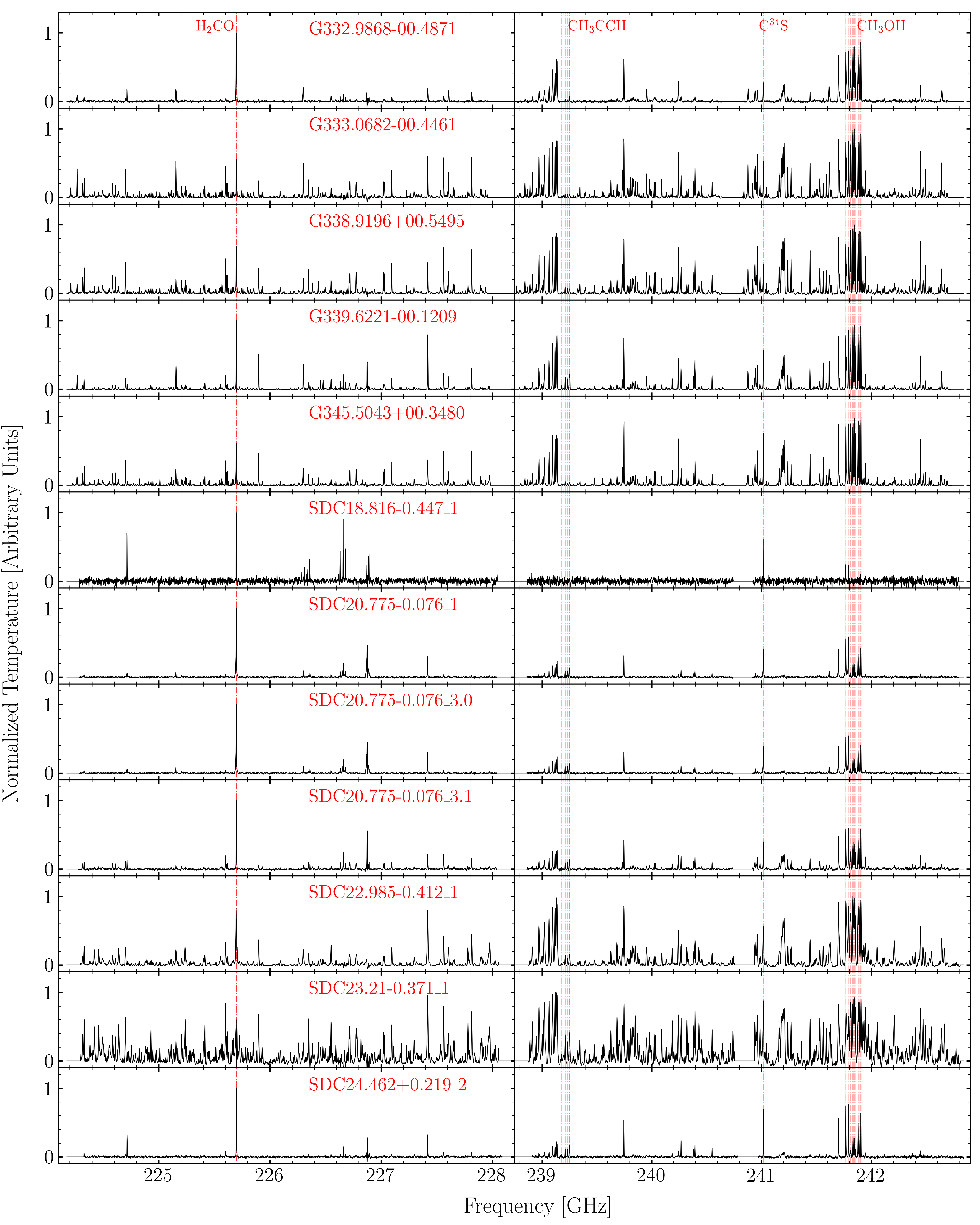}
        \repeatcaption{fig:norm_spectra_sample2}{continued.}
    \end{figure*}
    %
    \begin{figure*}
     \centering
             \includegraphics[width=\linewidth]{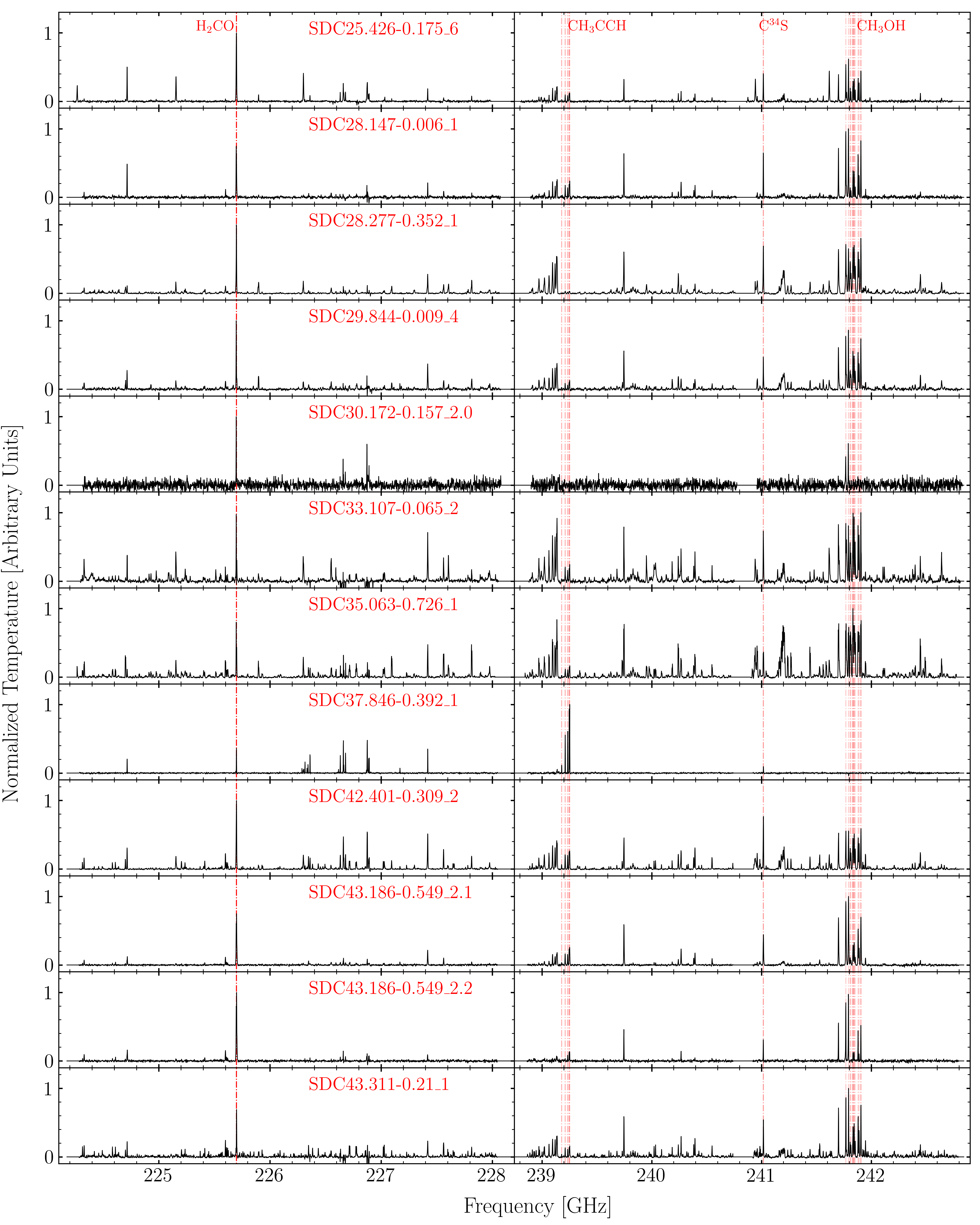}
        \repeatcaption{fig:norm_spectra_sample2}{continued.}
    \end{figure*}
\section{COM parameters availability} \label{appendix:COM_params}
    Table~\ref{tab:COMs_param} shows the availability of parameters from the rotational diagrams for each COM and object in the sample. Refer to the description in Sect.~\ref{subsec:COMs_selected}.
    %
    \begin{table*}[htb]
    \caption{\label{tab:COMs_param}Availability of parameters obtained from the COM rotational diagram analysis.}
    \centering
    \setlength{\tabcolsep}{12pt}
    \begin{tabular}{l c c c c} 
        \hline\hline
            \multicolumn{1}{c}{Source} & \multicolumn{1}{c}{\ch{CH3OH}} & \multicolumn{1}{c}{\ch{CH3CN}} & \multicolumn{1}{c}{\ch{CH3CCH}} & \multicolumn{1}{c}{LLE+GMM groups}\\
            \multicolumn{1}{c}{(1)} & (2) & (3) & (4) & (5) \\
            \hline
            G013.6562-00.5997 & \checkmark & \checkmark & \checkmark & 3 \\
            G017.6380+00.1566 & \checkmark & \checkmark & \checkmark & 2 \\
            G023.3891+00.1851 & \checkmark & \checkmark & \checkmark & 2 \\
            G029.8620-00.0444 & \checkmark & \checkmark & \checkmark & 2 \\
            G030.1981-00.1691 & \checkmark & \checkmark & \checkmark & 2 \\
            G034.7569+00.0247 & \checkmark & \checkmark & \checkmark & 2 \\
            G034.8211+00.3519 & \checkmark & \checkmark & \checkmark & 2 \\
            G050.2213-00.6063 & \checkmark & \checkmark \tablefootmark{$\ast$} & \checkmark \tablefootmark{$\ast$} & 1 \\
            G326.6618+00.5207 & \checkmark & \checkmark \tablefootmark{$\ast$} & \checkmark & 1 \\
            G327.1192+00.5103 & \checkmark & \checkmark & \checkmark & 3 \\
            G332.0939-00.4206 & \checkmark & \checkmark & \checkmark & 3 \\
            G332.9636-00.6800 & \checkmark & \checkmark & \checkmark & 2 \\
            G332.9868-00.4871 & \checkmark & \checkmark & \checkmark & 2 \\
            G333.0682-00.4461 & \checkmark & \checkmark & \checkmark & 3 \\
            G338.9196+00.5495 & \checkmark & \checkmark & \checkmark & 3 \\
            G339.6221-00.1209 & \checkmark & \checkmark & \checkmark & 2 \\
            G345.5043+00.3480 & \checkmark & \checkmark & \checkmark & 2 \\
            SDC18.816-0.447\_1 & \checkmark & \checkmark \tablefootmark{$\ast$} & \checkmark \tablefootmark{$\ast$} & 1 \\
            SDC20.775-0.076\_1 & \checkmark & \checkmark \tablefootmark{$\ast$} & \checkmark & 1 \\
            SDC20.775-0.076\_3.0 & \checkmark & \checkmark & \checkmark & 2 \\
            SDC20.775-0.076\_3.1 & \checkmark & \checkmark & \checkmark & 2 \\
            SDC22.985-0.412\_1 & \checkmark & \checkmark & \checkmark & 3 \\
            SDC23.21-0.371\_1 & \checkmark & \checkmark & \checkmark & 3 \\
            SDC24.381-0.21\_3\tablefootmark{+} & \checkmark \tablefootmark{$\ast$} & \checkmark \tablefootmark{$\ast$} & \checkmark & ... \\
            SDC24.462+0.219\_2\tablefootmark{a} & \tablefootmark{$\ast\ast\ast$} & \checkmark & \checkmark & P2 (91\%)\tablefootmark{\dag} \\
            SDC25.426-0.175\_6 & \checkmark & \checkmark & \checkmark & 2 \\
            SDC28.147-0.006\_1\tablefootmark{a} & \tablefootmark{$\ast\ast\ast$} & \checkmark & \checkmark & P2 (91\%)\tablefootmark{\dag} \\
            SDC28.277-0.352\_1 & \checkmark & \checkmark & \checkmark & 2 \\
            SDC29.844-0.009\_4\tablefootmark{a} & \tablefootmark{$\ast\ast\ast$} & \checkmark & \checkmark & P2 (91\%)\tablefootmark{\dag} \\
            SDC30.172-0.157\_2.0 & \checkmark & \checkmark \tablefootmark{$\ast$} & \checkmark \tablefootmark{$\ast$} & 1 \\
            SDC30.172-0.157\_2.1\tablefootmark{+} & ... & ... & ... & ...  \\
            SDC33.107-0.065\_2 & \checkmark & \checkmark & \checkmark & 2 \\
            SDC35.063-0.726\_1\tablefootmark{a} & \tablefootmark{$\ast\ast\ast$} & \checkmark & \checkmark & P3 (53\%)\tablefootmark{\dag} \\
            SDC37.846-0.392\_1\tablefootmark{a} & \checkmark \tablefootmark{$\ast$} & \checkmark \tablefootmark{$\ast$} & \checkmark & P1 (61\%)\tablefootmark{\dag} \\
            SDC42.401-0.309\_2 & \checkmark & \checkmark & \checkmark & 2 \\
            SDC43.186-0.549\_2.1\tablefootmark{a} & \tablefootmark{$\ast\ast\ast$} & \checkmark & \checkmark & P2 (87\%)\tablefootmark{\dag} \\
            SDC43.186-0.549\_2.2 & \checkmark & \checkmark & \checkmark & 1 \\
            SDC43.311-0.21\_1 & \checkmark & \checkmark & \checkmark & 2 \\
            SDC43.877-0.755\_1 & \checkmark & \checkmark & \checkmark & 2 \\
            SDC45.787-0.335\_1 & \checkmark & \checkmark & \checkmark & 2 \\
            SDC45.927-0.375\_2.0\tablefootmark{a} & \tablefootmark{$\ast\ast\ast$} & \checkmark & \checkmark & P1 (65\%)\tablefootmark{\dag} \\
            SDC45.927-0.375\_2.1\tablefootmark{a} & ... & ... & ...  & P1 (61\%)\tablefootmark{\dag} \\
            SDC45.927-0.375\_2.2\tablefootmark{a} & \tablefootmark{$\ast\ast\ast$} & ... & ... & P1 (65\%)\tablefootmark{\dag} \\  
        \hline
    \end{tabular}
    \tablefoot{
    All parameters are taken from \citet{Frimpong_2021}. Columns: (1) name of the sources. (2), (3), and (4) parameters (temperature and column density) of \ch{CH3OH}, \ch{CH3CN}, and \ch{CH3CCH}, respectively, measured by \citet{Frimpong_2021}. (5) Unsupervised classification groups obtained from the LLE+GMM method. The values of the parameters are detailed in Table 3.2 of \cite{Frimpong_2021} and \cite{Frimpong_2023subm}.\\
    \tablefoottext{$\ast$}{Sources with assumed molecular temperatures. The column densities are upper limits.\\}
    \tablefoottext{$\ast\ast\ast$}{Spectra with two components of \ch{CH3OH} in emission.\\}
    \tablefoottext{+}{Spectrum not considered for PCA due to a lack of \ch{CH3OH}, \ch{CH3CN}, and \ch{CH3CCH} emission.\\}
    \tablefoottext{a}{Spectrum not considered in the LLE and GMM analysis due to the lack of physical and chemical parameters of the COMs studied here.\\}
    \tablefoottext{\dag}{Supervised classification predicted by the RF plus PCA analysis. P1, P2, and P3 correspond to groups 1, 2, and 3. The prediction probability is shown in parentheses.\\}
    }
    \end{table*}
\end{appendix}
    
\end{document}